\documentclass[utf8]{frontiersSCNS} 

\usepackage{url,hyperref,lineno,microtype,subcaption}
\usepackage[onehalfspacing]{setspace}

\def\keyFont{\fontsize{8}{11}\helveticabold }
\def\firstAuthorLast{Das {et~al.}} 
\def\Authors{Ankan Das\,$^{1,*}$, Milan Sil\,$^1$, Rana Ghosh\,$^{1}$, Prasanta Gorai\,$^2$, Soutan Adak\,$^3$, Subhankar Samanta\,$^3$, Sandip K. Chakrabarti\,$^1$}
\def\Address{$^1$Indian Centre for Space Physics, 43 Chalantika, Garia Station Road, Kolkata 700084, India \\
$^2$Department of Space, Earth \& Environment, Chalmers University of Technology, SE-412 96 Gothenburg, Sweden \\
$^3$Ramakrishna Mission Residential College, Narendrapur, Kolkata 700103, India}
\def\corrAuthor{Ankan Das}

\def\corrEmail{ankan.das@gmail.com}

\begin{document}
\onecolumn
\firstpage{1}

\title[Encounter desorption]{Effect of binding energies on the encounter desorption} 

\author[\firstAuthorLast ]{\Authors} 
\address{} 
\correspondance{} 

\extraAuth{}

\maketitle
 \begin{abstract}
 The abundance of interstellar ice constituents is usually expressed with respect to the water ice because, in denser regions, a significant portion of the interstellar grain surface would be covered by water ice. The binding energy (BE), or adsorption energy of the interstellar species regulates the chemical complexity of the interstellar grain mantle. 
Due to the high abundance of water ice, the BE of surface species with the water is usually provided and widely used in astrochemical modeling. However, the hydrogen molecules would cover some part of the grain mantle in the denser and colder part of the interstellar medium. Even at around $\sim 10$ K, few atoms and simple molecules with lower adsorption energies can migrate through the surface. 
The BE of the surface species with H$_2$ substrate would be very different from that of a water substrate.  However, adequate information regarding these differences is lacking.  Here, we employ the quantum chemical calculation to provide the BE of 95 interstellar species with H$_2$ substrate. These are representative of the BEs of species to a H$_2$ overlayer on a grain surface. On average, we notice that the BE with the H$_2$ monomer substrate is almost ten times lower than the BE of these species reported earlier with the H$_2$O c-tetramer configuration. The encounter desorption of H and H$_2$ was introduced (with $\rm{E_D(H,H_2)=}45$ K and $\rm{E_D(H_2,H_2)=}23$ K) to have a realistic estimation of the abundances of the surface species in the colder and denser region.  Our quantum chemical calculations yield higher adsorption energy of H$_2$ than that of H ($\rm{E_D(H,H_2)=}23 - 25$ K and $\rm{E_D(H_2,H_2)=}67 - 79$ K). 
We further implement an astrochemical model to study the effect of encounter desorption with the present realistic estimation. The encounter desorption of the N atom (calculations yield $\rm{E_D(N,H_2)=}83$ K) is introduced to study the differences with its inclusion.
\tiny
 \keyFont{ \section{Keywords:} astrochemistry, binding energy, numerical, ISM, star formation, chemical model, Monte Carlo (MC) algorithm} 

 \end{abstract}

\section{Introduction} \label{sec:intro}
Interstellar grains mainly consist of amorphous silicate and some form of carbonaceous materials \citep{li04}. It is now well established that these grains can significantly constrain the chemical composition of molecular clouds or star-forming regions. In the cloud's denser regions, where the temperature is reasonably low ($\sim 10$ K), the grain surface would be covered by icy layers. A sizable portion of these icy layers may contain water molecules. Thus, providing the binding energies (BEs) with the water as a substrate is proper. In reality, the surface species would face a bare grain in the diffuse region. Some ice layers would grow on the top of this grain surface in the denser medium and host the incoming species. The composition of this ice layer depends on the initial elemental abundance of the species in that region. It would not necessarily always be H$_2$O dominated. There are ample examples of the presence of a notable portion of CO, CO$_2$, CH$_4$, NH$_3$, CH$_3$OH, etc., on the ice \citep{gibb04,das10,das11,das16,gora20a}.

Hydrogen molecules are ubiquitous in denser regions of the interstellar medium (ISM). Thus, its accretion rate on the grain is much higher in comparison to the others.
However, because of the low adsorption energy, it can be easily desorbed from the grain. Despite this, a significant portion of the grain mantle would be covered by molecular hydrogen, especially around the cold and dense interstellar condition. This inhomogeneous surface coverage could influence the mobility of the other surface species. Initially, the encounter desorption mechanism was introduced by \cite{hinc15} to eliminate the overestimation of the abundance of molecular hydrogen on the grain. 
This desorption process occurs during surface diffusion and is induced by the presence of repulsive inter-H$_2$ forces, effectively reducing the BE of H$_2$.
They considered  $\rm{gH_2+gH_2 \rightarrow H_2 + gH_2}$, where, ``g" designates the grain surface species. They obtained an excellent match within the microscopic Monte Carlo method and the rate equation approach when they implemented this unique approach. 
The Monte Carlo approach is best suited for monitoring the chemical composition of the grain mantle. However, it is time-consuming \citep{chak06a,chak06b,das08a,das10,das11,das16,cupp07}.  
Recently, \cite{chang20} considered a similar process and included H's desorption by a similar mechanism. They considered $\rm{gH+gH_2 \rightarrow H + gH_2}$, which means, whenever the surface H meets one surface H$_2$, surface H desorb with a certain probability. They reported a significant difference between the formation of some key surface species with the inclusion of this treatment. 

A substantial amount of BE values are available from the temperature-programmed desorption (TPD) studies on various model substrates like graphite, diamond-like carbon, amorphous or crystalline silica, silicates, water, and other ice surfaces \citep{coll04,ward12,nobl12,duli13}. But, the BE of the species with H$_2$ substrate is yet to be known. \cite{cupp07} had estimated the BE of H atom on H$_2$ substrate $\sim 45$ K by following
\cite{vida91}. They also estimated the BEs of O, OH, $\rm{H_2,\ O_2,\ H_2O,\ O_3,\ O_2H,\ and \ H_2O_2}$ with the H$_2$ substrate by scaling its obtained BE with H$_2$O substrate with the ratio of BE between the BE of H with water substrate and with H$_2$ substrate.

A vital impediment in examining the encounter desorption with other species is the shortage of information about the adsorption energy of these species with H$_2$ molecule. Here, we employ quantum chemical calculations to determine the adsorption energy of these species with H$_2$ molecule. Obtained BE assessments are executed in our Chemical Model of Molecular Cloud (hereafter CMMC) \citep{das15a,das15b,gora17a,gora17b,sil18,gora20b}. The encounter desorption effect is vital during the prestellar core phase.
Study the formation of stars is one of the essential intricacies of astrophysics. A complete understanding of the star formation process is yet to be established. However, in brief,  stars are formed by a long condensation process \citep{paga12}. In the beginning, warm diffuse material ($\sim 8000$ K) converts into a cold neutral atomic gas ($\sim 100$ K and $\sim 10 - 100$ cm$^{-3}$). After further evolution, it transforms into a more dense region ($10^2-10^4$ cm$^{-3}$ and $\sim 10-20$ K). If no other heating source is present, then a dense core ($> 10^4$ cm$^{-3}$) appears in some places of these turbulent materials. Some of these cores further evolve into prestellar cores ($> 10^5$ cm$^{-3}$) \citep{berg07,keto08}. Prestellar cores further continue their evolution for the formation of the protostar. Due to the accretion of atoms and molecules, gas-phase abundance is depleting, whereas the molecular ice mantles form. The chemical composition of the grain mantle is mainly governed by the addition of atomic hydrogen with the atoms or simple molecules. The chemical composition of the bulk ices further varies with the star formation process associated with it. Depending on this, it is expected that the ice composition would be very different in various places. However, from the infrared observations, it was observed that the significant repositories of interstellar hydrogen, oxygen, carbon, and nitrogen are H$_2$O, CH$_3$OH, H$_2$CO, CO, CO$_2$, CH$_4$, and NH$_3$ \citep{gibb00,whit07,ober08,boog15}.

This paper is compiled as follows. In Section \ref{sec:comp}, we confer computational methodology. Discussion and results are presented in Section \ref{sec:results}, and finally, in Section \ref{sec:conclusions}, we conclude.

\begin{table}
\tiny
\caption{Calculated binding energy (with MP2/aug-cc-pVDZ) of various species with H$_2$ monomer surface.}
\vskip 0.2cm
\begin{tabular}{cccccccccc}
\hline
{\bf Sl.}& {\bf Species} & {\bf Ground} & \multicolumn{2}{c}{\bf Binding Energy}& {\bf Sl.} & {\bf Species} & {\bf Ground} & \multicolumn{2}{c}{\bf Binding Energy}\\
\cline{4-5}
\cline{9-10}
 {\bf No.} & & {\bf State} & {\bf in K} & {\bf in kJ/mol} & {\bf No.} && {\bf State} & {\bf in K} & {\bf in kJ/mol} \\
\hline
1 & H & doublet & 23 (25$^a$), 45$^c$ & 0.189 (0.210$^a$) &51&CO$_2$& singlet &241 & 2.003 \\
2 & H$_2$ & singlet & 67 (79$^a$), 23$^c$, 100$^d$ & 0.549 (0.659$^a$) & 52&OCS& singlet &257& 2.137 \\
3&He& singlet &27& 0.226 &53&SO$_2$& singlet &324& 2.691 \\
4&C& triplet &50 & 0.417 & 54&CH$_3$& doublet &198& 1.644 \\
5&N& quartet &83 (78$^a$)& 0.690 (0.651$^a$) & 55&NH$_3$& singlet &455 & 3.781 \\
6&O& triplet &46, 55$^c$ & 0.386 & 56 & SiH$_3$ & doublet & 159 & 1.321 \\
7&Na& doublet &22& 0.184 &  57 & C$_2$H$_2$ & singlet & 337 & 2.799 \\
8&Mg& singlet &62& 0.514 & 58 & N$_2$H$_2$ & singlet & 608 & 5.059 \\
9&Si& triplet &642& 5.343 &  59 & H$_2$O$_2$ & singlet & 628, 340$^c$ & 5.222 \\
10&P& quartet &107 & 0.887 & 60 & H$_2$S$_2$ & singlet & 573 & 4.763 \\
11&S& triplet & 88 & 0.732 & 61 & H$_2$CN & doublet & 376 & 3.130 \\
12&NH& triplet &286& 2.381 & 62 & H$_2$CO &singlet& 507 & 4.219 \\ 
13&OH& doublet &380, 240$^c$ & 3.158 &  63 & HC$_2$N&triplet &413 & 3.434 \\
14&PH& triplet &151& 1.258 & 64&HC$_2$O&doublet&326 & 2.712 \\
15&C$_2$& triplet &204& 1.696 & 65 & HNCO &singlet& 289 & 2.405 \\
16&HF& singlet &287& 2.386 & 66 & H$_2$CS&singlet&545 & 4.532 \\
17&HCl& singlet &162& 1.350 & 67&C$_3$O& singlet & 414 & 3.442 \\
18&CN& doublet &4695& 39.041 & 68&CH$_4$&singlet&138 & 1.150 \\
19&N$_2$& singlet &198 & 1.649 & 69&SiH$_4$&singlet&165 & 1.370 \\
20 & CO & singlet &215& 1.788 & 70&C$_2$H$_3$& doublet &265 & 2.200 \\
21 & SiH & doublet & 188 & 1.562 & 71 & CHNH$_2$ & singlet & 858 / 463$^b$ & 7.133 / 3.846$^b$ \\
22 & NO & doublet & 159 & 1.321 &72 & CH$_2$NH& singlet& 602 & 5.007 \\
23 &O$_2$& triplet &159, 69$^c$ & 1.321 & 73 &c-C$_3$H$_2$& singlet & 472 & 3.925 \\
24&HS& doublet &222&1.848 & 74&CH$_2$CN&doublet&440 & 3.662 \\
25&SiC& triplet &212& 1.759 &75&CH$_2$CO&singlet&276 & 2.297 \\
26&CP& doublet &165&1.373&76&HCOOH&singlet&369 & 3.066 \\
27&CS& singlet &337&2.804 & 77 & CH$_2$OH & doublet & 272 & 2.263 \\
28&NS& doublet & 353 / 171$^b$ & 2.938 / 1.423$^b$ & 78 & NH$_2$OH & singlet & 2770 & 23.028 \\
29&SO& triplet &337&2.801 & 79 & HC$_3$N & singlet & 427 & 3.555 \\
30&S$_2$& triplet &187&1.552 &  80&C$_5$& singlet &379 & 3.156 \\
31&CH$_2$& triplet &165&1.376 &  81&C$_2$H$_4$&singlet&250 & 2.079 \\
32&NH$_2$& doublet &347& 2.888 & 82&CH$_2$NH$_2$& doublet &428 & 3.560 \\
33&H$_2$O& singlet &360, 390$^c$& 2.993 & 83&CH$_3$OH&singlet & 414 / 258$^b$ & 3.445 / 2.145$^b$ \\
34&PH$_2$& doublet &178& 1.483 & 84 & CH$_2$CCH & doublet & 105 & 0.872 \\
35&C$_2$H& doublet & 242 & 2.014 & 85&CH$_3$CN& singlet &453 & 3.765 \\
36&N$_2$H& doublet &432 & 3.589 & 86&CH$_3$NH$_2$& singlet & 610 & 5.072 \\
37&O$_2$H& doublet &339, 300$^c$ & 2.819 & 87&C$_2$H$_5$& doublet &327 & 2.720 \\
38&HS$_2$& doublet &660 &5.487 & 88&CH$_3$CCH& singlet &125 & 1.040 \\
39&HCN & singlet &395 &3.282 &89&CH$_2$CCH$_2$& singlet &489 & 4.070 \\
40&HNC & singlet &338 & 2.814 &90&CH$_3$CHO& singlet &573 & 4.765 \\
41&HCO & doublet &243 & 2.019 &91&PN& singlet &399 & 3.324 \\
42&HOC & doublet & 769 & 6.396 & 92 &PO& doublet &509 & 4.230  \\
43&HCS & doublet & 334 & 2.780 & 93 & SiN & doublet & 154 & 1.281 \\
44&HNO & singlet &574 & 4.773 &94&F& doublet &24 & 0.202 \\
45&H$_2$S & singlet &99 & 0.824 &95&C$_2$H$_5$OH& singlet & 590 & 4.906 \\
46&C$_3$& singlet &295 & 2.455 &&&& & \\
47&O$_3$& singlet & 381, 120$^c$ & 3.169 & &&&&\\
48&C$_2$N& doublet &339 & 2.817 & &&&&\\
49&C$_2$S& triplet & 355 & 2.951 & &&&&\\
50&OCN& doublet & 422 & 3.510 & &&&&\\
\hline
\end{tabular} \\
\vskip 0.2cm
$^a$ The BE values for the adsorbates H, H$_2$, and N with the adsorbent as H$_2$ considering IEFPCM model are noted in parentheses. \\
$^b$ Alternative BE values are for different binding sites.\\
$^c$ \cite{cupp07}. \\
$^d$ \cite{sand93}.
\label{tab:BE}
\end{table}

\section{Computational details} \label{sec:comp}
\subsection{Quantum chemical calculations}
Here, we have utilized the Gaussian 09 suite of programs \citep{fris13} for quantum chemical calculations. In a periodic treatment of surface adsorption phenomena, the BE is related to the interaction energy ($\Delta$E), as:
\begin{equation}
    BE = -\Delta E
\end{equation}
For a bounded adsorbate, the BE is a positive quantity and is defined as:
\begin{equation}
    BE = (E_{surface}+E_{species})-E_{ss},
\end{equation}
where $\rm{E_{ss}}$ is the optimized energy for the complex system where a species is placed at a suitable distance from the grain surface. $\rm{E_{surface}}$ and $\rm{E_{species}}$ are the optimized energies of
the grain surface and species, respectively.

To find the optimized energy of all structures, we have used a Second-order M\o ller-Plesset (MP2) method with an aug-cc-pVDZ basis set \citep{dunn89}. 
We have considered $95$ interstellar species for the computation of their BEs with the H$_2$ substrate.
To make the calculation more straightforward, we have considered a monomer configuration of the H$_2$ molecule as an adsorbent.  The adsorbates noted in Table \ref{tab:BE} are placed at a suitable
distance from the adsorbent with a weak bond so that a Van der Waals interaction occurs during the optimization. All the optimized geometries are provided in the supplementary information. We must confess that the interstellar species considered in this study are often larger than the H$_2$. Since the estimated BEs are different in different locations, this may lead to a fallacious estimation. It is recommended to take the average whenever different binding sites are found. Despite these flaws, it can provide us with a general picture and startup initiative to compare the BE of a species with the water and H$_2$ substrate.
Following the BE calculations carried out by \cite{das18} (see Tables 2 and 3), here, we have not considered the ZPE and BSSE corrections for our BE calculations.
All the obtained BE values are noted in Table \ref{tab:BE}. It is interesting to note that except for phosphorous, the calculated adsorption energy of most of the abundant atoms (H, C, N, O, and S) with the H$_2$ substrate is found to be $<100$ K.

In Table \ref{tab:BE}, we have reported the BE values obtained by considering a free-standing H$_2$ interacting with a species. But in reality, this H$_2$ would be pre-adsorbed and can feel the surface. It would yield a different BE than the previous case.
To check the effect of condensed H$_2$O in the ice phase, we also have calculated the interaction energy by considering the molecule embedded in a continuum solvation field. For this purpose, we have examined the local effects and the integral equation formalism (IEF) variant of the polarizable continuum model (PCM) \citep{canc97,toma05} with water as a solvent \citep{gora20a}. The obtained values for H, H$_2$, N with the IEFPCM model are noted in Table \ref{tab:BE} (in parentheses). The two calculations significantly differ. For example, with the free-standing H$_2$, we have the BE of H, H$_2$, and N $\sim$ 23 K, 67 K, and 83 K, respectively, whereas with the IEFPCM model, we have obtained $\sim$ 25 K, 79 K, and 78 K, respectively. So, the free-standing H$_2$ underestimated the BE of H and H$_2$ by 2 K and 12 K, whereas it overestimated N's case by 5 K. We also provided in Table \ref{tab:BE} the literature BE values \citep{cupp07,sand93} (if available) for the comparison.

\cite{das18} provided BEs of the $\sim 100$ interstellar species considering the c-tetramer configuration of water molecules. Table \ref{tab:BE} shows the BE of the roughly same interstellar species with the H$_2$ monomer. We have noticed that the obtained BEs with H$_2$ are much smaller than those of the water tetramer configuration. On average, we have received almost ten times lower BEs with the H$_2$ surface. Table \ref{tab:BE} shows the ground state of the species used to calculate the BE. The values of BE are very much sensitive on the chosen ground state spin multiplicity. To evaluate the ground state spin multiplicity of each species, we have taken the help of Gaussian 09 suite of program. The way to check for the ground state spin multiplicity is to run separate calculations (job type ``opt+freq"), each with different spin multiplicities, and then compare the results between them. The lowest energy electronic state solution of the chosen spin multiplicity is the ground state noted for the species in Table \ref{tab:BE}.  \\

\subsection{Astrochemical model}
We have included the encounter desorption phenomenon in our CMMC code \citep{das15a,das15b,das16,gora17a,gora17b,sil18,gora20b} to study its effect. The surface chemistry network of our model is mostly adopted from \cite{ruau15,das15b,gora20b}. The gas-phase network of the CMMC model is mainly adopted from the UMIST database \citep{mcel13}. Additionally, we have also included the deuterated gas-phase chemical network from the UMIST.
A cosmic ray rate of $1.3 \times 10^{-17}$ s$^{-1}$ is considered in all our models. 
Cosmic ray-induced desorption and non-thermal desorption rate with a fiducial parameter
of $0.01$ is considered. For all the grain surface species, we have adopted a photodesorption rate of
$1 \times 10^{-4}$ per incident UV photon \citep{ruau15}.
A sticking coefficient of $1.0$ is considered for the neutral species except for the H and H$_2$. The sticking coefficients of H and H$_2$ are considered by following the relation proposed by \cite{chaa12}.
Following \cite{garr11}, here, we have implemented the competition between diffusion, desorption, and reaction. 
For the diffusion energy ($E_b$), we have considered R$\times$ adsorption energy ($E_D$). Here, R is a scaling factor that can vary between 0.35 and 0.8 \citep{garr07}.
The BE of the species is mostly considered from \cite{wake17}, and a few from \cite{das18}. Table \ref{tab:init} refers to the adopted initial abundances concerning the total hydrogen nuclei in all forms. Except for HD's value in Table \ref{tab:init}, elemental abundances are taken from \cite{seme10}. We considered the initial abundances of HD from \cite{robe00}.  

  \begin{table}
  \scriptsize
  \centering
    \caption{Initial elemental abundances considered in this study.}
    \vskip 0.5 cm
    \begin{tabular}{cc}
    \hline
    {\bf Species} & {\bf Abundances} \\
    \hline
        H$_2$ & $5.00 \times 10^{-1}$ \\
        He & $9.00 \times 10^{-2}$\\
        N& $7.60 \times 10^{-5}$\\
        O& $2.56 \times 10^{-4}$\\
        C$^+$&$1.20 \times 10^{-4}$ \\
        S$^+$& $8.00 \times 10^{-8}$\\
        Si$^+$ &$8.00 \times 10^{-9}$\\
        Fe$^+$& $3.00 \times 10^{-9}$\\
        Na$^+$& $2.00 \times 10^{-9}$\\
        Mg$^+$&$7.00 \times 10^{-9}$\\
        Cl$^+$&$2.00 \times 10^{-10}$\\
        HD&$1.60 \times 10^{-5}$\\
        \hline
    \end{tabular}
    \label{tab:init}
\end{table}

The encounter desorption effect was first introduced by \cite{hinc15}. The rate of encounter desorption of H$_2$ on the surface of H$_2$ is defined as:
\begin{equation}
    En_{H_2}= \frac{1}{2}\ k_{H_2,H_2} \ gH_2\  gH_2\ P(H_2,H_2),
    \label{eqn:enH2}
\end{equation}
where gH$_2$ is the surface concentration of H$_2$ molecules in cm$^{-3}$, $\rm{P(H_2,H_2)}$ defines the probability of desorption over the diffusion, and
$\rm{k_{H_2,H_2}}$ is the diffusion rate coefficient over the H$_2$O substrate. 
$\rm{k_{H_2,H_2}}$ is defined as follows \citep{hase92}: 
\begin{equation}
k_{H_2,H_2}= \kappa (R_{diffH_2} + R_{diffH_2})/n_d \ \ cm^3s^{-1}.
\end{equation}
In the above equation, $n_d$ is the dust-grain number density, $\kappa$ is the probability for the reaction to happen (unity for the exothermic reaction without activation energy), and $R_{diff}$ is the diffusion of the species.  $\rm{P(H_2,H_2)}$ in Equation \ref{eqn:enH2} is defined as:
\begin{equation}
    {\rm P(H_2,H_2)=\frac{Desorption \ rate \  of \ H_2 \ on \ H_2 \ substrate}{Desorption \ rate \ of \ H_2 \ on \ H_2 \ substrate \ + \ Diffusion \ of \ H_2 \ on \ H_2 \ substrate}}.
    \label{eqn:prob}
\end{equation}
There would be various desorption factors (by thermal, reactive, cosmic ray etc.). The thermal desorption is defined as: $\nu \ {\rm exp(-E_D(H_2,H_2)/T)}$ s$^{-1}$, where T is the dust temperature. Similarly, there would be various diffusion mechanisms, but thermal diffusion would be the dominating. It is defined as: $ \rm{\nu \ exp(-E_b(H_2,H_2)/T)/S\ (s^{-1}) = \ thermal \ hopping \ rate/ number \ of \ sites} \ (s^{-1})$. Here, we have used $\rm{Diffusion\ energy\ (E_b)\ =\ R \ \times Adsorption\ energy\ (E_D)}$.
Recently, \cite{chang20} has extended this work by considering the encounter desorption of the H atom. In their definition of the encounter desorption of H$_2$, in Equation \ref{eqn:prob}, they used the hopping rate of H$_2$ on H$_2$ substrate instead of the diffusion rate of H$_2$ on H$_2$ substrate. Following the prescription defined in \cite{chang20}, the encounter desorption of species X is defined as:
\begin{equation}
 {\rm En_{X,H_2} \ = \ \frac{h_{X,H_2}}{S} gX \ gH_2 \ P(X,H_2) \ P_X},
\end{equation}
where $\rm{h_{X,H_2}}$ is the hopping rate over H$_2$O surface ($\nu \ exp(-\rm{E_b(X,H_2)/T})$), P(X,H$_2$) is the desorption probability of gX while encountering with gH$_2$, and P$_X$ denotes the probability of gX to migrate at the location of gH$_2$ over the H$_2$O substrate. P(X,H$_2$) and P$_X$ are defined as,
\begin{equation}
        {\rm P(X,H_2)=\frac{Desorption \ rate \  of \ X \ on \ H_2 \ substrate}{Desorption \ rate \ of \ X \ on \ H_2 \ substrate \ + \ Hopping\ rate \ of \ X \ on \ H_2 \ substrate}},
        \label{eqn:chan}
        \end{equation}
            \begin{equation}
    P_X =\frac{\rm Hopping \ rate \ of \ X \ on \ H_2O \ substrate}{\rm Hopping \ rate\  of \ X \ on \ H_2O \ substrate \ + \ Hopping \ rate \ of \ H_2 \ on \ H_2O \ substrate}.
  \end{equation}

  \begin{figure}
    \centering
    \includegraphics[height=10cm,width=8cm,angle=-90]{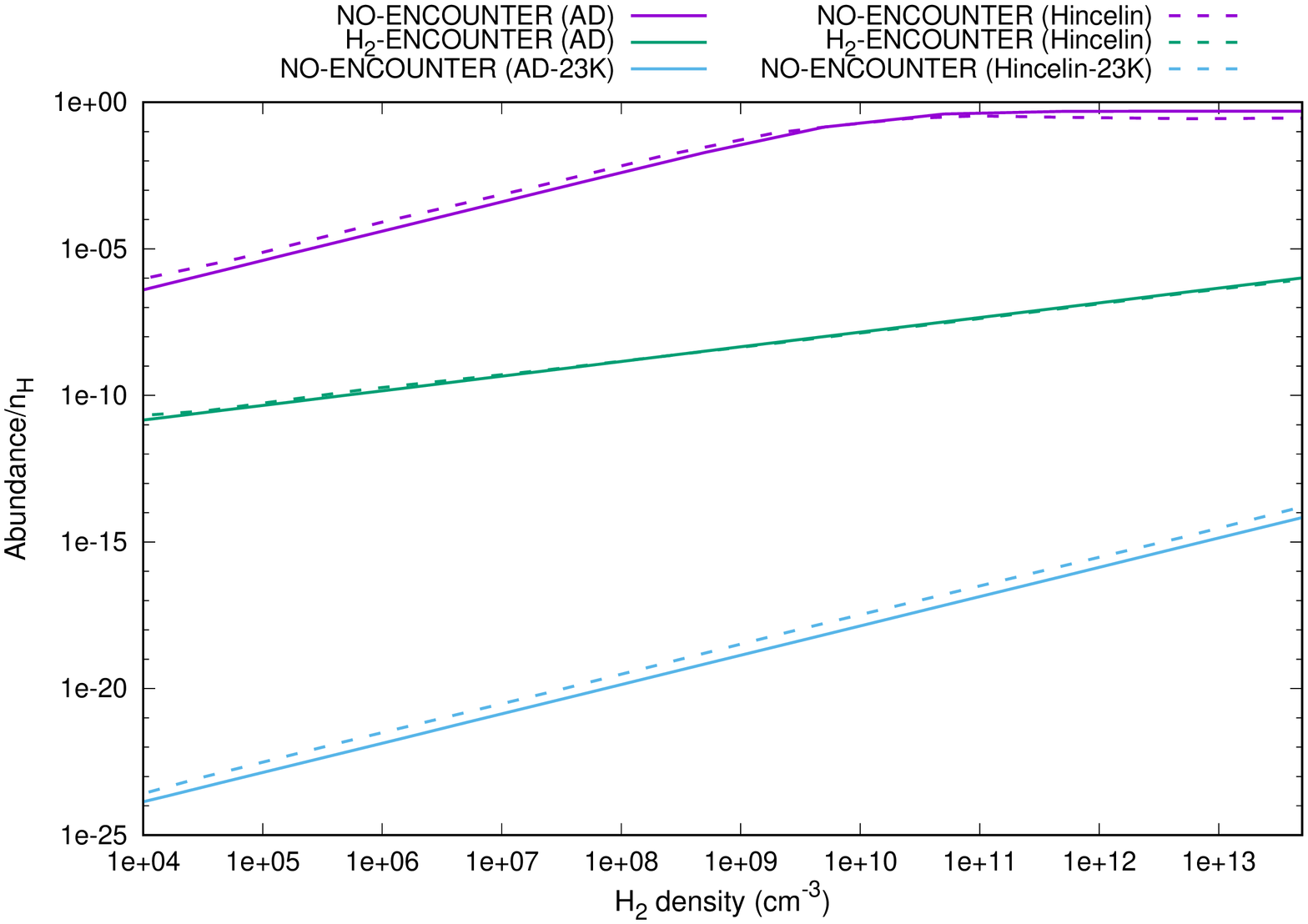}
    \caption{The comparison between Figure 2 of \cite{hinc15} and the cases obtained here. We have extracted Figure 2 of \cite{hinc15} by using the online tool of \cite{roha20}. Three cases are shown: a) no encounter desorption is considered with $\rm{E_D(H_2,H_2O)=440}$ K, (b) no encounter desorption is considered with $\rm{E_D(H_2,H_2)=23}$ K, (c) encounter desorption of H$_2$ was considered with $\rm{E_D(H_2,H_2O)=440}$ K and $\rm{E_D(H_2,H_2)=23}$ K. We have noticed an excellent match between our calculated (solid curves) steady-state abundance of H$_2$ on grain surface and that obtained in \cite{hinc15} (dashed curves).}
    \label{fig:comp}
\end{figure}
  
  \begin{table}
\tiny
    \caption{The obtained abundance of gH$_2$, gH, gH$_2$O and gCH$_3$OH for the effect of encounter desorption of H$_2$ under various situation with $R=0.35$, $n_H=10^7$ cm$^{-3}$, and $T=10$ K.}
   \hskip -1.5 cm
    \vskip 0.5 cm
    \begin{tabular}{cccccc}
    \hline
   {\bf Case} & {\bf Case specification} & \multicolumn{4}{c}{\bf Abundance at $10^6$ years with ${\rm n_H}=10^{7}$ cm$^{-3}$}\\
   \cline{3-6}
   {\bf No.} && {\bf gH$_2$} & {\bf gH} & {\bf gH$_2$O} & {\bf gCH$_3$OH} \\
   &&&& {\bf (\% increase)} &  {\bf (\% increase)} \\
    \hline
    \multicolumn{6}{c}{$\rm{E_D(H,H_2O})=450$ K}\\
    \hline
        1& No encounter desorption &$  2.0011749 \times 10^{-4}$&$2.311976  \times 10^{-24}$&$  8.8375039 \times 10^{-5}$&$ 8.8416179 \times 10^{-6}$\\ 
        &&&& (0.00) & (0.00) \\
        2& ${\rm E_D(H_2,H_2)=23}$ K & $1.183836 \times 10^{-11}$& $ 3.2799079 \times 10^{-24}$&$ 8.927643 \times 10^{-5}$ &$  6.3629209 \times 10^{-6}$\\
        &\citep{hinc15}&&& ($1.02$) & ($-28.03$) \\
         3& ${\rm E_D(H_2,H_2)=23}$ K &$2.7660509 \times 10^{-11}$&$3.253898 \times 10^{-24}$&$8.9327989 \times 10^{-5}$ & $ 6.4358829 \times 10^{-6}$\\
         &\citep{chang20}&&& (1.08) & ($-27.21$) \\
         4& ${\rm E_D(H_2,H_2)=67}$ K & $1.051303 \times 10^{-10}$&$2.25004 \times 10^{-24}$&$1.02707 \times 10^{-4}$ & $6.119676 \times 10^{-6}$\\
         &\citep{chang20}&&& (16.22) & ($-30.79$) \\
         5& ${\rm E_D(H_2,H_2)=79}$ K & $  1.5474109 \times 10^{-10}$&$  2.2527589 \times 10^{-24}$&$ 1.028505 \times 10^{-4}$ & $6.2066129 \times 10^{-6}$\\
         &\citep{chang20}&&& (16.38) & ($-29.8$) \\
         \hline
        \multicolumn{6}{c}{$\rm{E_D(H,H_2O)=650}$ K}\\
        \hline
        6& No encounter desorption & $2.00117 \times 10^{-4}$&$ 1.467684 \times 10^{-21}$&$9.434889 \times 10^{-5}$&$ 5.799469 \times 10^{-6}$\\
        &&&& (0.00) & (0.00) \\
         7& ${\rm E_D(H_2,H_2)=67}$ K &$1.051293 \times 10^{-10}$&$2.0551489 \times 10^{-21}$&$ 9.361905 \times 10^{-5}$ &$4.559358 \times 10^{-6}$\\
         &\citep{chang20}&&& ($-0.77$) & ($-29.80$) \\
         \hline
    \end{tabular}
    \label{tab:H2abunnew}
\end{table}

  \begin{figure}
    \centering
    \includegraphics[height=18cm,width=15cm,angle=-90]{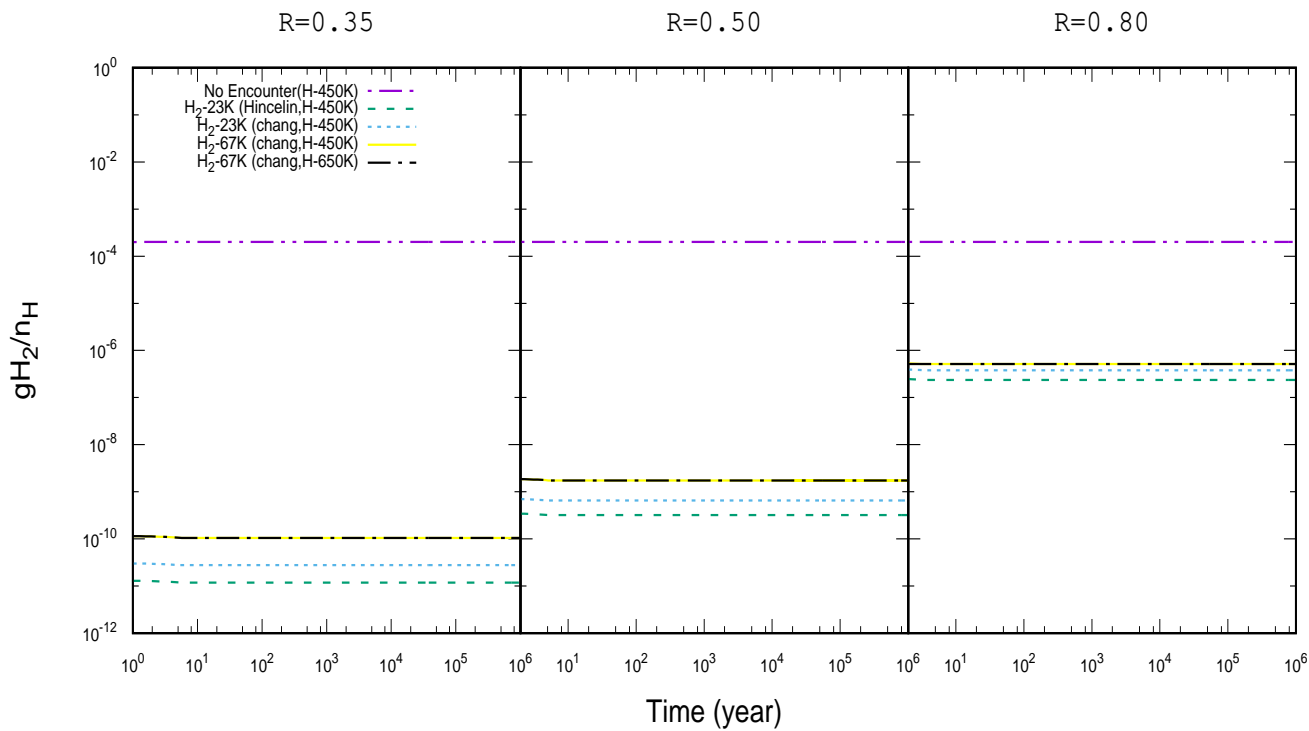}
    \vskip -2 cm
    \caption{Time evolution of the abundances of gH$_2$ with $\rm{n_H=10^7}$ cm$^{-3}$ and T$=10$ K are shown for $R=0.35$, $0.5$, and $0.8$. The dash-dotted purple curve represents the time evolution of gH$_2$ abundance with the no encounter desorption (with $\rm{E_D(H,H_2O) = 450 \ K}$). It depicts that the gH$_2$ abundance remains roughly invariant with the changes in R. However when encounter desorption is introduced, gH$_2$ abundance increases with the $R$. The time evolution of the gH$_2$ abundance with $\rm{E_D(H_2,H_2) =23}$ K and $\rm{E_D(H,H_20) = 450}$ K is shown with the green dashed line when the method of \cite{hinc15} is used and blue dotted line when the method of \cite{chang20} is used. gH$_2$ abundances obtained with our estimated BE value (i.e., $\rm{ E_D(H_2,H_2)= 67}$ K) are shown with a solid yellow line. For this case, we have used ${\rm E_D(H,H_2O)=450}$ K and the method used in \cite{chang20}. With the black dash-dotted line, the time evolution of gH$_2$ abundance is shown with ${\rm E_D(H,H_2O)=650}$ K and method of \cite{chang20}. We have seen significant differences when we have used different energy barriers and different methods \citep{hinc15,chang20}. Obtained values of gH$_2$ are further noted in Table \ref{tab:H2abunnew} for better understanding.}
    \label{fig:H2-abun}
\end{figure}

\begin{figure}
    \centering
    \includegraphics[height=18cm,width=15cm,angle=-90]{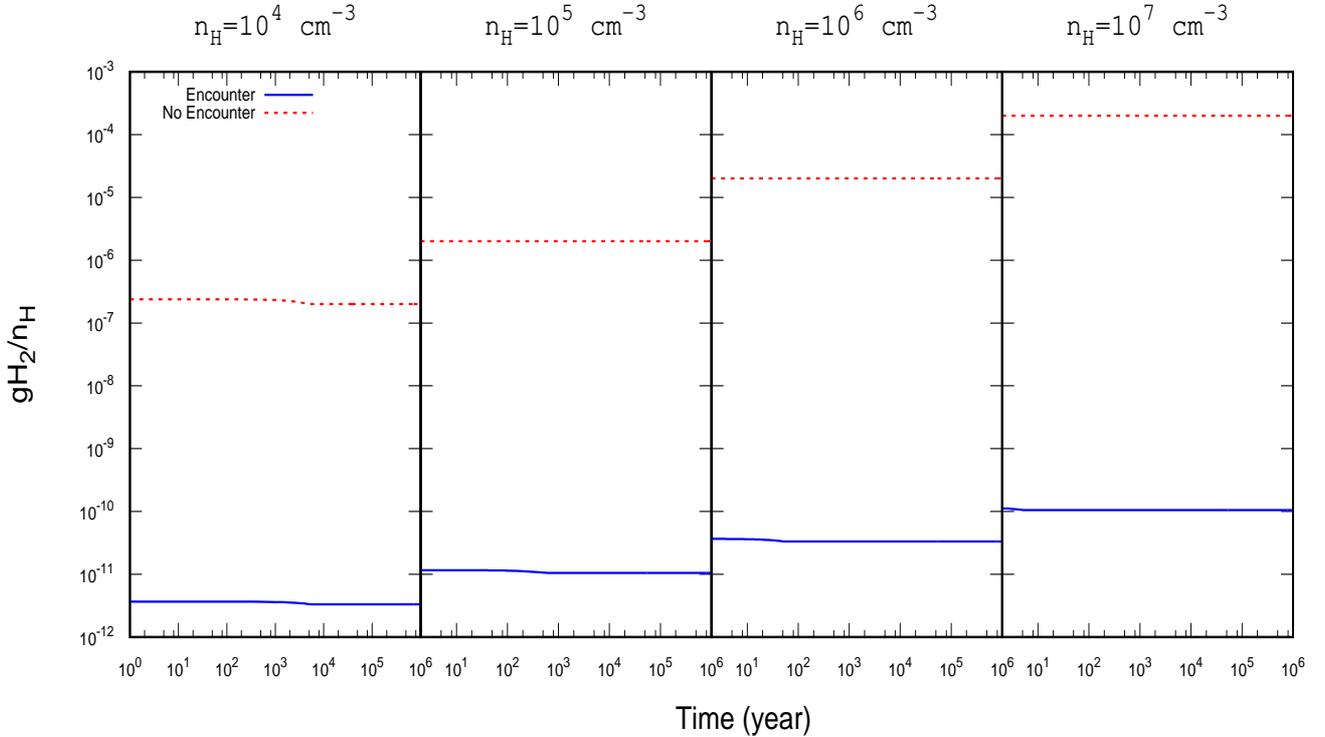}
    \vskip -2cm
    \caption{Time evolution of gH$_2$ with $R=0.35$ and various $\rm{n_H}$ ($10^4, \ 10^5,$ $10^6$, and $10^7$ cm$^{-3}$) are shown. It depicts that the effect of encounter desorption increases with the increase in density.}
    \label{fig:H2-abun-den}
\end{figure}

\begin{figure}
    \centering
    \includegraphics[height=18cm,width=15cm,angle=-90]{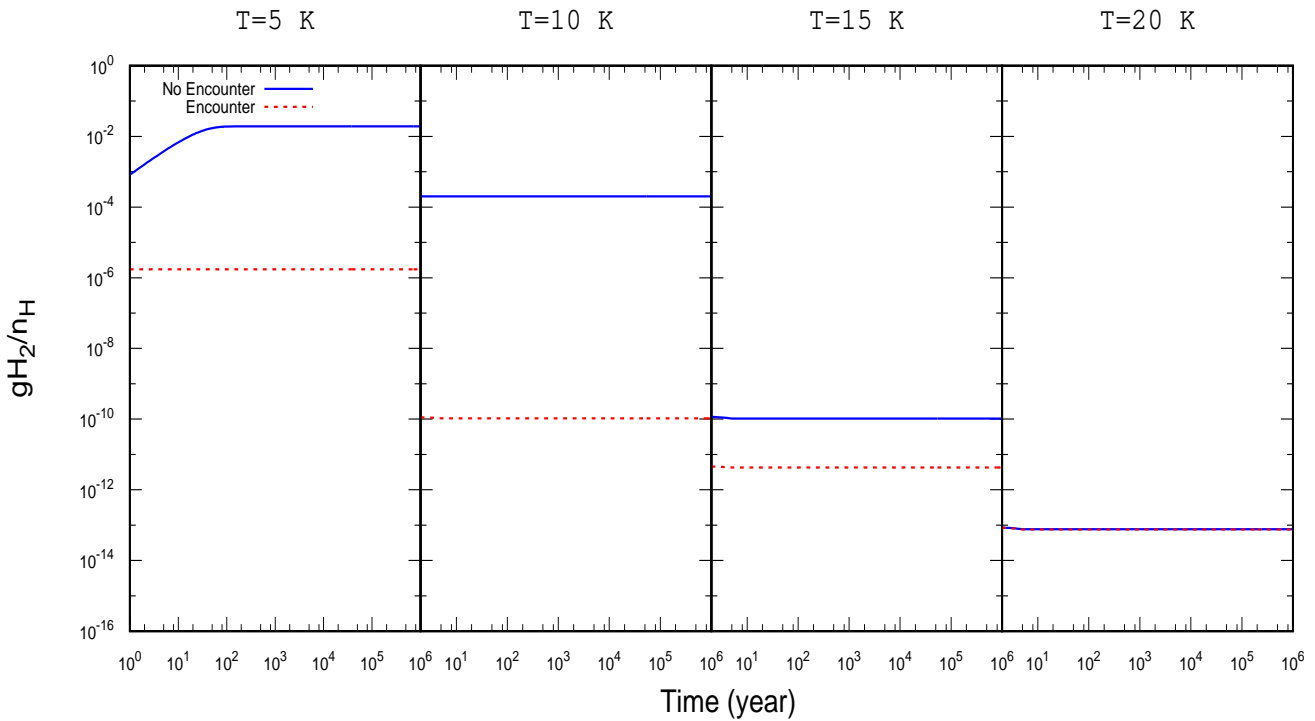}
    \vskip -2cm
    \caption{Time evolution of gH$_2$ with R$=0.35$, $\rm{n_H=10^7}$ cm$^{-3}$ and various temperatures (5, 10, 15, and 20 K) are shown. It depicts that the effect of encounter desorption decreases with the increase in temperature.}
    \label{fig:H2-abun-temp}
\end{figure}

\begin{figure}
    \centering
    \includegraphics[height=5cm,width=5cm,angle=-90]{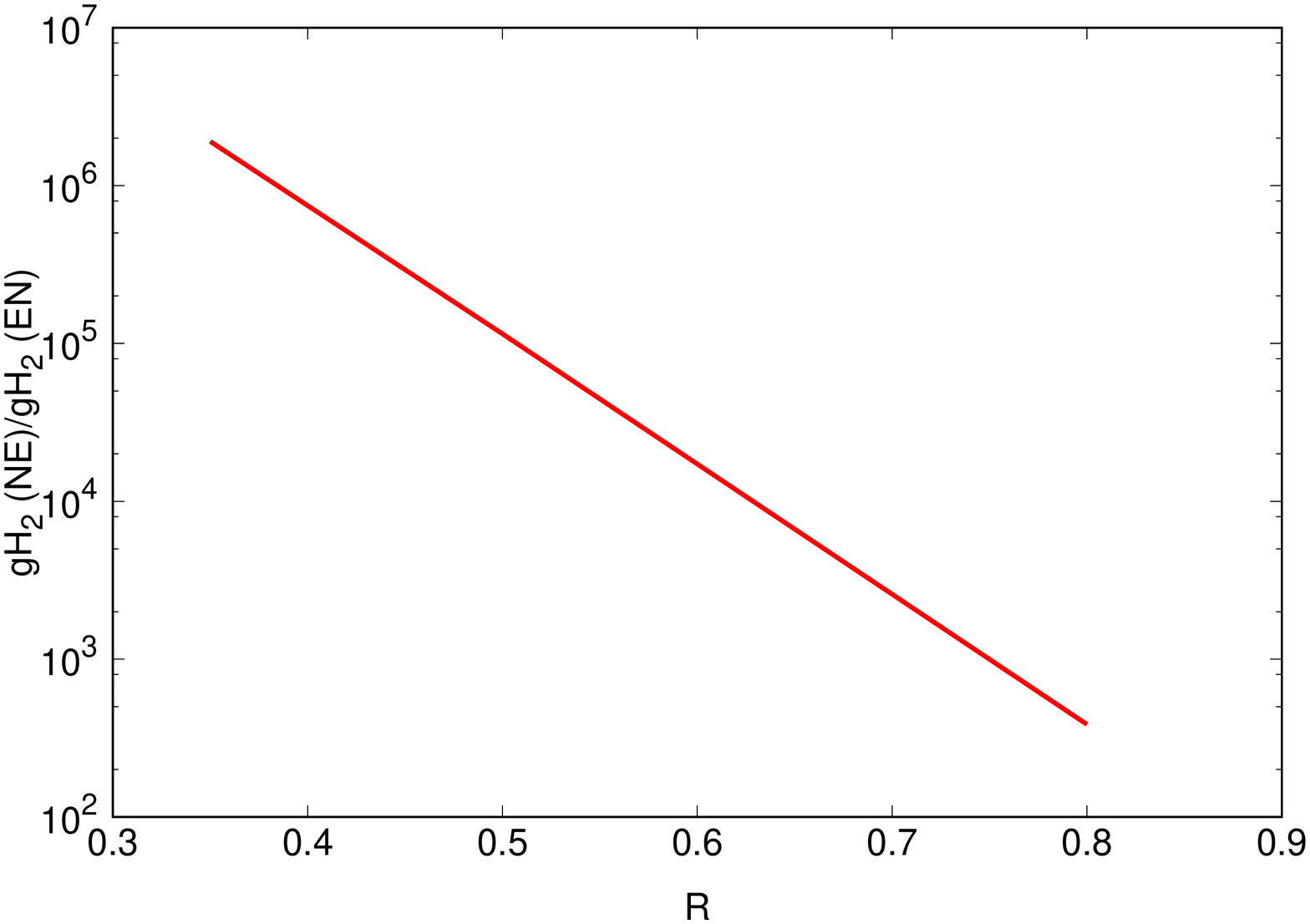}
     \includegraphics[height=5cm,width=5cm,angle=-90]{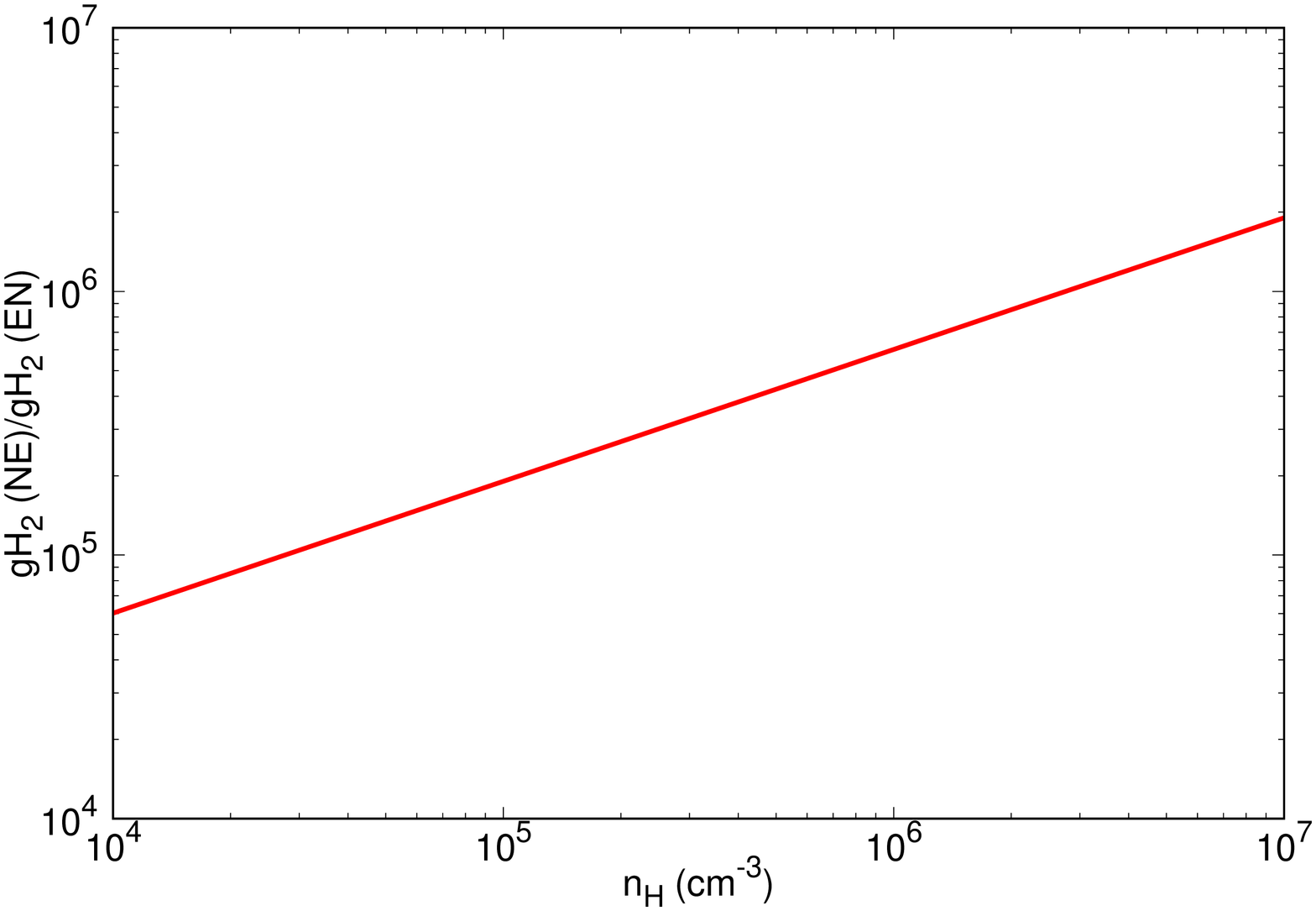}
      \includegraphics[height=5cm,width=5cm,angle=-90]{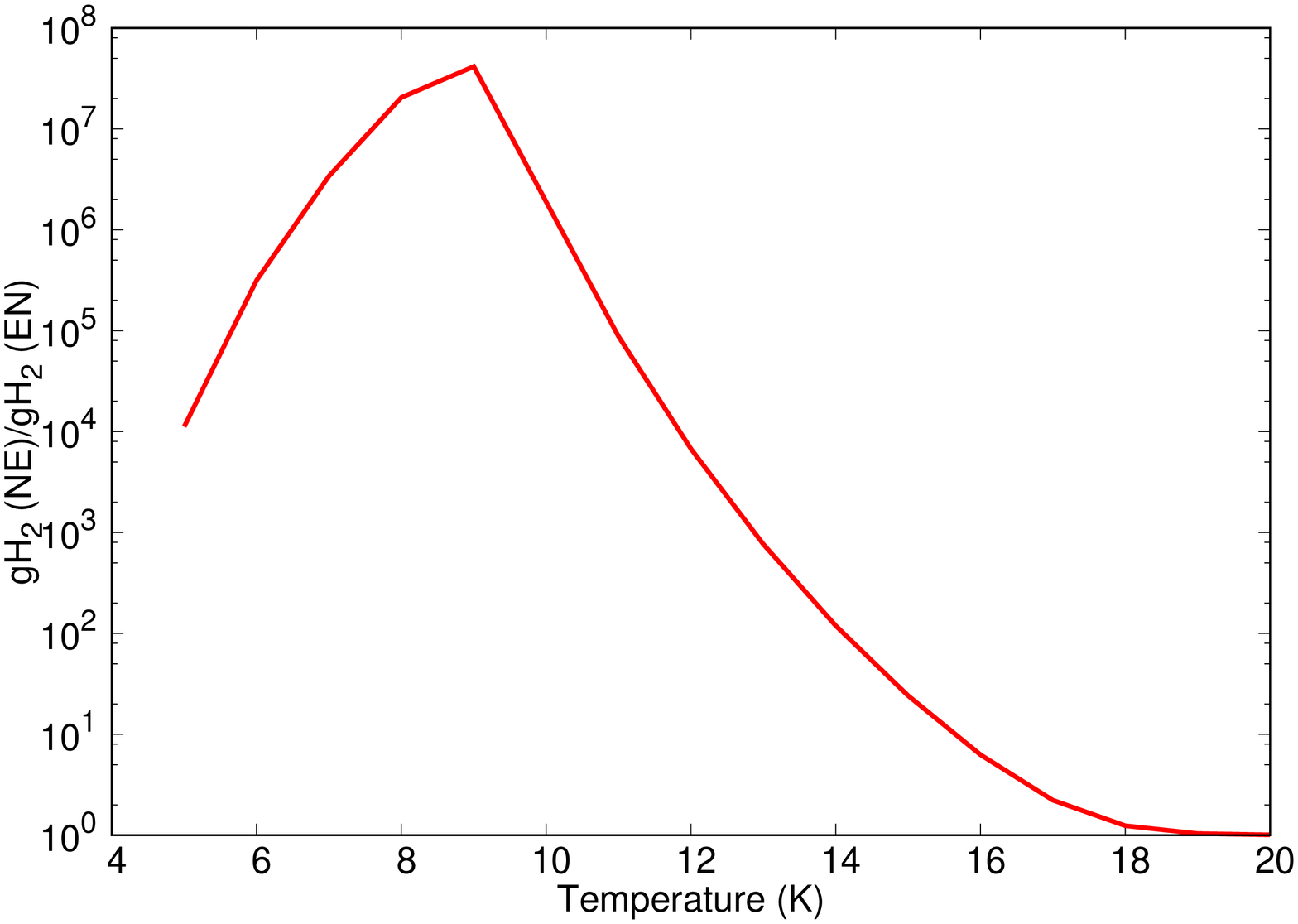}
    \caption{The ratio between the final abundances of gH$_2$ obtained with the no encounter (NE) desorption and encounter desorption (EN) is shown. From left to right, it shows the variation of this ratio with R, $\rm{n_H}$, and temperature, respectively.}
    \label{fig:H2-ratio}
\end{figure}

\section{Results and Discussion} \label{sec:results}
\subsection{Encounter desorption of H$_2$ \label{sec:H2-EN}}
First of all, we have benchmarked our model with \cite{hinc15}. In Figure \ref{fig:comp}, we have compared our results with those obtained in \cite{hinc15}. 
For this comparison, following \cite{hinc15}, we have used T$=10$ K, $\rm{E_D(H_2,H_2O)= 440}$ K, $\rm{E_D(H,H_2O)=450}$ K, $\rm{E_D(H_2,H_2)}=23$ K, and R=0.5. 
Solid curves in Figure \ref{fig:comp} represent the cases obtained here, and the rest are extracted from \cite{hinc15} by using the online tool of \cite{roha20}.
Our results with and without encounter desorption show an excellent match with \cite{hinc15}. Presently in the KIDA database (\url{kida.astrophy.u-bordeaux.fr}), more updated BE values were listed. It suggests that  $\rm{E_D(H,H_2O)=650}$ K. The results obtained from our quantum chemical calculations shown in Table \ref{tab:BE} represent the estimated BE values with the H$_2$ substrate. In the following section, we have used these updated energy values, and the effects of their changes are discussed. \\

\subsubsection{gH$_2$}
Figure \ref{fig:H2-abun} shows the time evolution of gH$_2$ by considering $n_H=10^7$ cm$^{-3}$, $T=10$ K and $R = 0.35-0.80$.  
Interestingly the abundance of gH$_2$ seems to be invariant with R's changes, whereas it strongly depends on R in encounter desorption.  R's lower value means a quicker hopping rate, whereas a  higher value represents a delayed hopping rate.
With the increase in R, gH$_2$ abundance raises for the encounter desorption case. It means that as we rise R's value, the encounter desorption effect depreciates.
The left panel of Figure \ref{fig:H2-ratio} exposes that with the increase in R's value, a steady decrease in the ratio between the gH$_2$ abundance with no encounter desorption case (NE) and with encounter desorption case (EN) is obtained. The probability of the encounter desorption is inversely proportional to the rate of diffusion (Equation \ref{eqn:prob}) or hopping (Equation \ref{eqn:chan}). Since the increase in the value of R induces faster diffusion and hopping, it is lowering the encounter desorption probability of H$_2$ as expected.
Figure \ref{fig:H2-abun-den} shows the time evolution of gH$_2$ with NE and EN when we have used $R=0.35$, $T=10$ K and $n_H = 10^4 - 10^7$ cm$^{-3}$. 
In both cases, abundances of gH$_2$ increase with the density. 
The middle panel of Figure \ref{fig:H2-ratio} shows that the gH$_2$ abundance ratio between NE and EN with density. It depicts that the effect of encounter desorption is more pronounced for the higher density.
Figure \ref{fig:H2-abun-temp} shows that the gH$_2$ abundances when we have used $n_H=10^7$, $R=0.35$, and T$=5-20$ K. In the right panel of Figure \ref{fig:H2-ratio}, we have shown the gH$_2$ abundance ratio obtained between NE and EN with the temperature changes.
From the figures, it is seen that the effect of encounter desorption is maximum toward the lower temperature ($\sim 10$ K), and it ceases around $20$ K. The curve is similar to the H$_2$ formation efficiency discussed in \cite{chak06a,chak06b} for olivine grain.  
With the decrease in temperature, H atoms' mobility decreases. Thus, the formation rate decreases. With the increase in temperature, the hopping rate increases, which can increase the formation efficiency, but at the same time, the residence time of H atoms decreases which affects the H$_2$ formation efficiency. As a result, the H$_2$ formation efficiency is maximum at around $\sim 10$ K, and the encounter desorption effect is pronounced at the peak hydrogen formation efficiency.

For a better illustration, the obtained abundances with $R=0.35$, T=10 K, and $\rm{n_H=10^7}$ cm$^{-3}$ are noted in Table \ref{tab:H2abunnew} at the end of the total simulation time ($\sim 10^6$ years). 
\cite{chang20} considered the competition between hopping rate and desorption rate of H$_2$ (Equation \ref{eqn:chan}), whereas \cite{hinc15} considered the battle between the diffusion and desorption rate of H$_2$ (Equation \ref{eqn:prob}). This difference in consideration resulting $\sim$ two times higher abundance of gH$_2$ with the consideration of \cite{chang20} compared to \cite{hinc15} (see case 2 and 3 of Table \ref{tab:H2abunnew} and Figure \ref{fig:H2-abun}).
Our quantum chemical calculation yields
$E_D(H_2,H_2)=67$ K, which is higher than it was used in the earlier literature value of $23$ K \citep{cupp07,hinc15,chang20}. The computed adsorption energy is further increased to $79$ K when we have considered the IEFPCM model. Table \ref{tab:H2abunnew} shows that increase in the BE ($E_D(H_2,H_2)=67$ K, and 79 K, case 4 and 5 of Table \ref{tab:H2abunnew}) results in sequentially higher surface coverage of gH$_2$ than it was with $E_D(H_2,H_2)=23$ K (case 3 of Table \ref{tab:H2abunnew}).
In case 5 of Table \ref{tab:H2abunnew}, we have noted the abundance of gH$_2$ when no encounter desorption effect is considered, but a higher adsorption energy of H atom is used ($E_D(H,H_2O)=650$ K). Case 6 of Table \ref{tab:H2abunnew} also considered this adsorption energy of H atom along with $E_D(H_2,H_2)=67$ K, and the method of \cite{chang20} is used. A comparison between the abundance of gH$_2$ of case 4 and case 6 (the difference between these two cases are in consideration of the adsorption energy of gH) yields a marginal decrease in the abundance of gH$_2$ when higher adsorption energy of gH is used. \\

\subsubsection{gH}
The obtained abundance of gH is noted in Table \ref{tab:H2abunnew}. The
gH abundance is marginally decreased in \cite{chang20} compared to \cite{hinc15}. The use of higher $\rm{E_D(H_2,H_2)}$ ($\sim 67$ K and $79$ K) lowers the value of gH compared to case 2. However, the use of the H atom's higher adsorption energy ($650$ K) can increase the gH abundance by a couple of orders of magnitude (see case 7 of Table \ref{tab:H2abunnew}). \\

\subsubsection{gH$_2$O,\ gCH$_3$OH}
The effect of the encounter desorption on the other major surface species (gH$_2$O and gCH$_3$OH) is also shown in Table \ref{tab:H2abunnew}. In the bracketed term, we have noted the percentage increase in their abundances from the case where no encounter desorption was considered (for $E_D(H,H_2O)=450$ K and $650$ K, respectively). 
Table \ref{tab:H2abunnew} depicts that the consideration of encounter desorption of H$_2$ can significantly change (decrease by $\sim 27-30\%$) the methanol abundance (case 3 and case 7) from that was obtained with the no encounter desorption (case 1 and case 6). However, the changes in the surface abundance of water are minimal ($\sim \pm 1\%$) for the addition of the encounter desorption of H$_2$. These changes (increase or decrease) are highly dependent on the adsorption energy of H, temperature, density, and the value of $R$ ($\sim 0.35$ noted in Table \ref{tab:H2abunnew}). 
The changes in $E_D(H_2,H_2)$ from $23$ K to $67$ K can influence the surface abundance of methanol and water. For example, in between case 3 and case 4 of Table \ref{tab:H2abunnew}, we can see that there is a significant increase ($\sim 15\%$) in the abundance of gH$_2$O when higher adsorption energy ($E_D(H_2,H_2)=67$ K) is used. However, this higher adsorption energy can marginally under-produce the methanol on the grain.
In brief, from Table \ref{tab:H2abunnew}, it is clear that the encounter desorption can significantly change the abundances of surface species. Still, these changes are highly dependent on the adopted adsorption energy with the water and H$_2$ ice and adopted physical parameters ($n_H$, $R$, $T$). \\

\begin{figure}
    \centering
    \includegraphics[height=12cm,width=10cm,angle=-90]{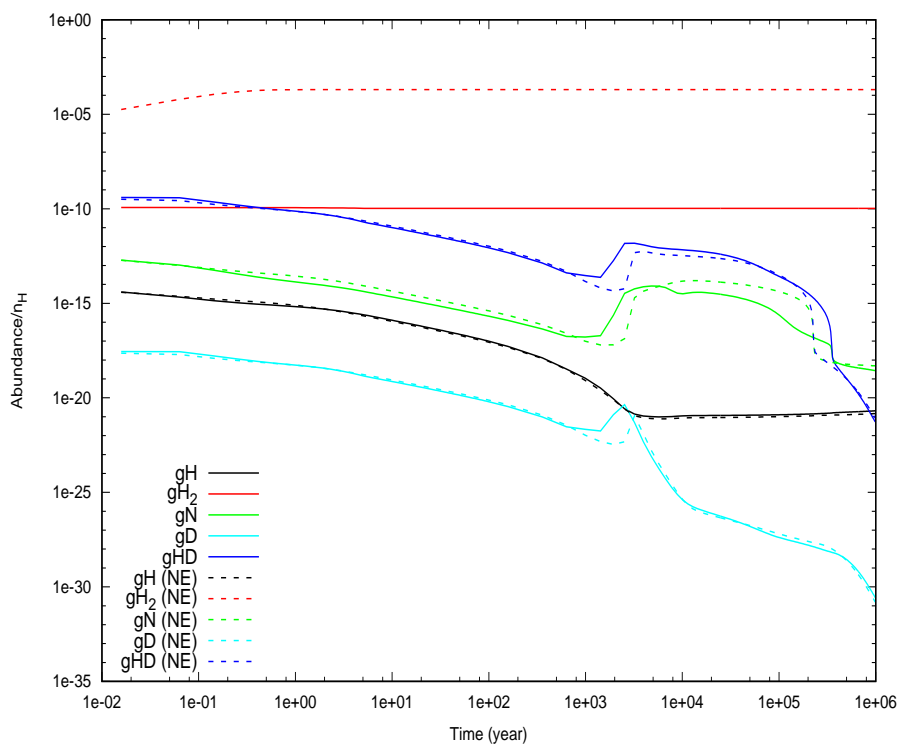}
    \caption{Time evolution of the abundances of H, H$_2$, D, HD, and N obtained from our simulation is shown. Solid curves represent the cases by considering the encounter desorption (with ${\rm E_D(H_2,H_2)=67}$ K) of H$_2$ and no encounter desorption of H$_2$ (dashed curves) with ${\rm E_D(H,H_2O)=650}$ K, $\rm{n_H=10^7}$ cm$^{-3}$, T$=10$ K, and $R=0.35$.}
    \label{fig:abun}
\end{figure}

\subsection{Encounter desorption of other species}
The idea of encounter desorption \citep{hinc15} primarily arose to eliminate the enhanced surface coverage of H$_2$ in the relatively denser and colder medium. Since H$_2$ has lower adsorption energy with the water surface ($\sim 440$ K), it could move on the surface very fast and occupy a position on the top of another H$_2$ molecule. Comparatively, in the denser and colder region, the chances of this occurrence enhance. Since the H$_2$ molecule on H$_2$ has negligible BE  \citep[23 K used in][]{cupp07,hinc15}, it could easily desorb back to the gas phase.
Other surface species can, of course, meet with H$_2$, but the idea of this encounter desorption arises when the species can occupy a position on the top of the H$_2$ molecule. For example, a carbon atom is having a BE of $10000$ K \citep{wake17}. H$_2$ could quickly meet one C atom on the grain surface, but due to the lower mobility of atomic carbon at a low temperature, every time H$_2$ will be on the top of the carbon atom. Since the whole C-H$_2$ system is attached to the water substrate; this will not satisfy the encounter desorption probability.
Among the various key elements considered in this study, gH, gN, and gF have the BE of 650 K \citep{wake17}, 720 K \citep{wake17}, and 800 K (listed in the original OSU gas-grain code from Eric Herbst group in 2006), respectively, with the water ice. It yields a reasonable timescale for hopping even at a low grain temperature ($\sim 10K$). Since the initial elemental abundance of F is negligible, we can neglect its contribution.
The hopping time scale is heavily dependent on the assumed value of $R$.
For example, by considering $R=0.35$, at 10 K, the hopping timescale for gH and gN is $1.12 \times 10^4$ years (with $\rm{E_D(H,H_2O)=650}$ K) and $4.61 \times 10^{-3}$ years (with $\rm{E_D(N,H_2O)=720}$ K), respectively. It changes to $1.9$ years and $226$ years for H and N atoms, respectively, for $R=0.5$.
Since the typical lifetime of a dark cloud is $\sim 10^6$ years, the criterion related to the encounter desorption is often satisfied. Among the di-atomic species, H$_2$ is only having a faster-swapping rate (having BE 440 K, which corresponds to a hopping time scale of $\sim 1.24 \times 10^{-7}$ years and $9 \times 10^{-5}$ years, respectively with R=0.35 and R=0.5). Looking at the faster hopping rate and their abundances on the grain surface, we have extended the consideration of the encounter desorption of these species. We have considered $\rm{gX+gH_2 \rightarrow X + gH_2}$, where X refers to H$_2$, H, and N.

\begin{figure}
    \centering
    \includegraphics[height=8cm,width=7cm,angle=-90]{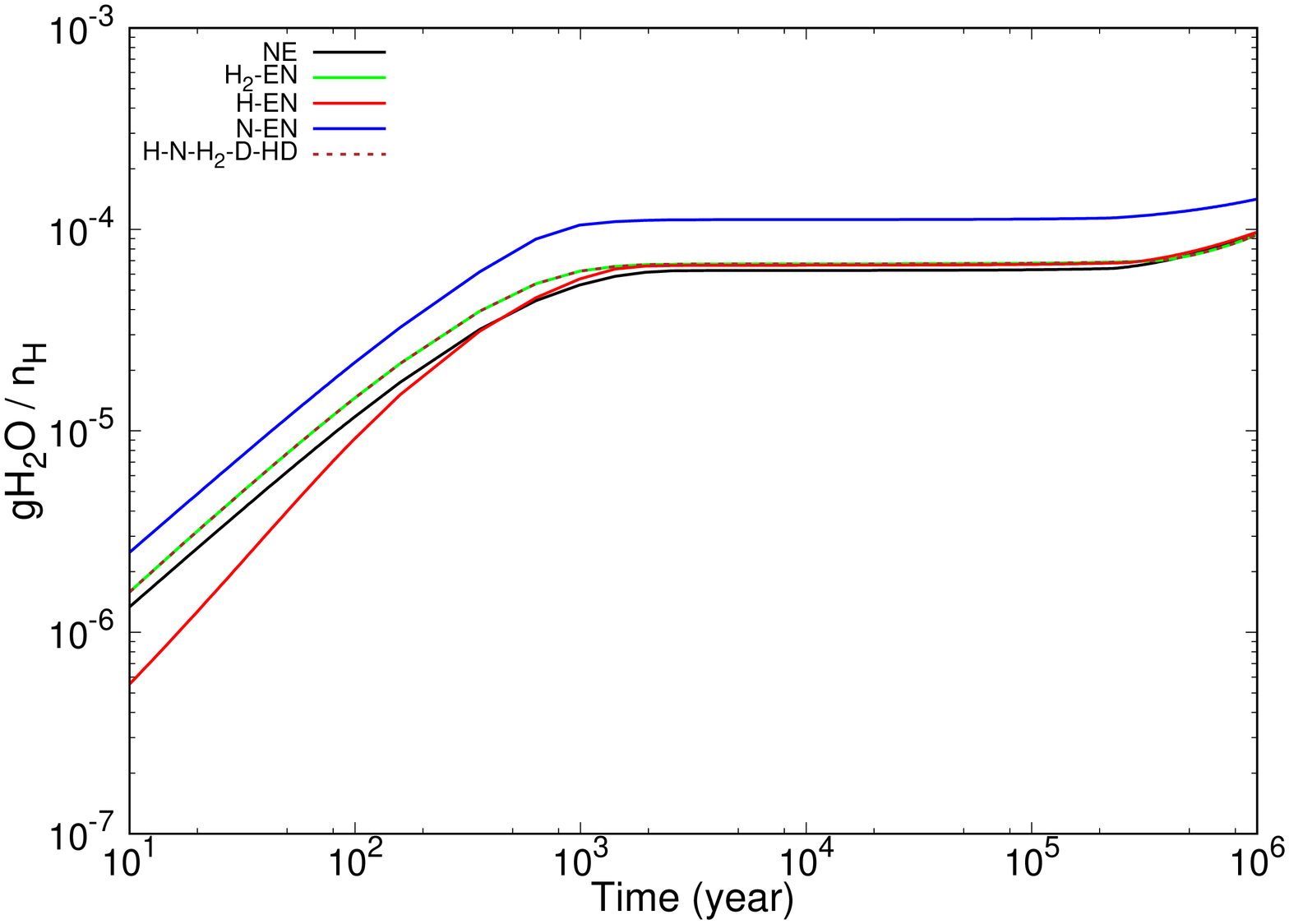}
    \includegraphics[height=8cm,width=7cm,angle=-90]{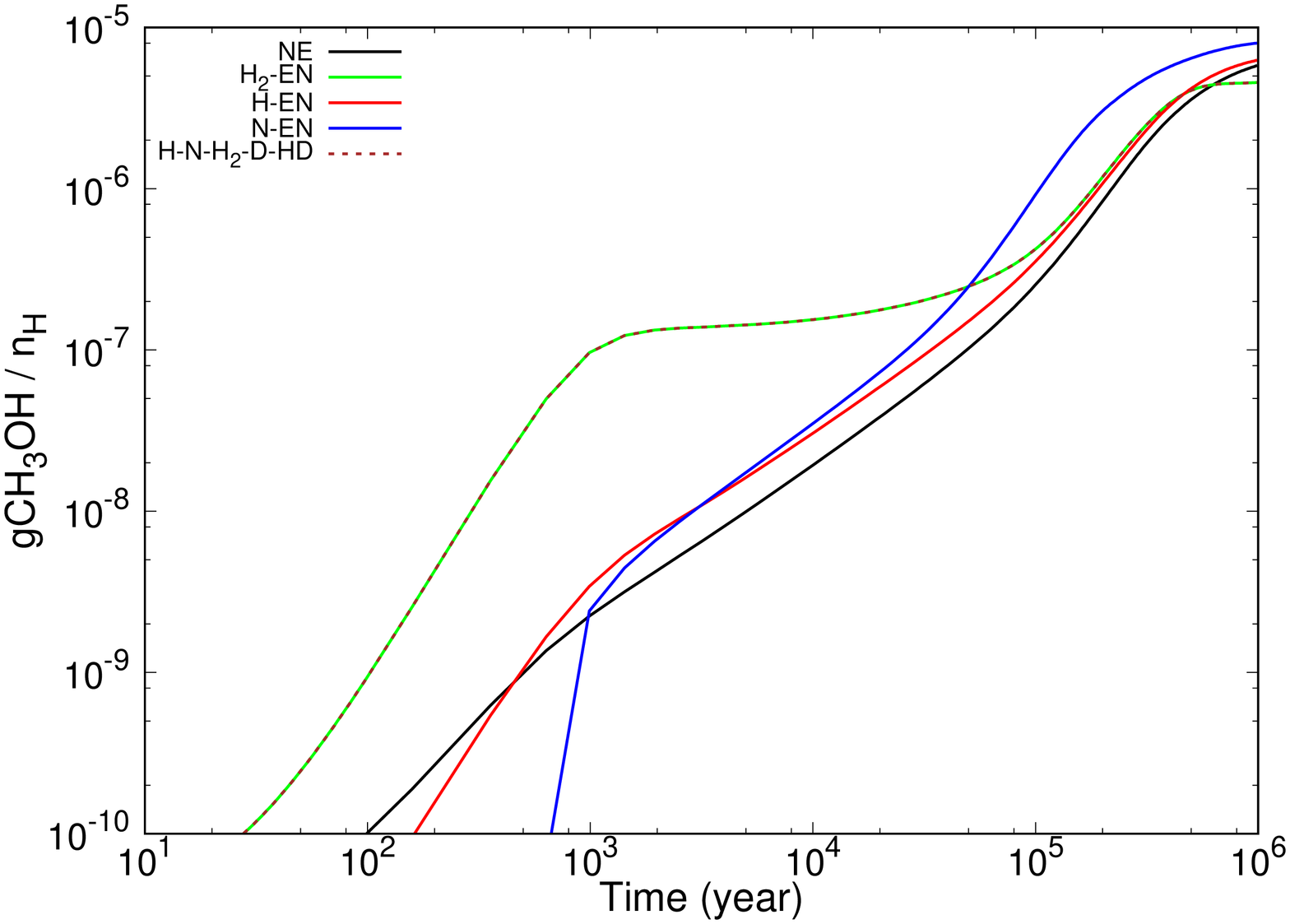}
    \includegraphics[height=8cm,width=7cm,angle=-90]{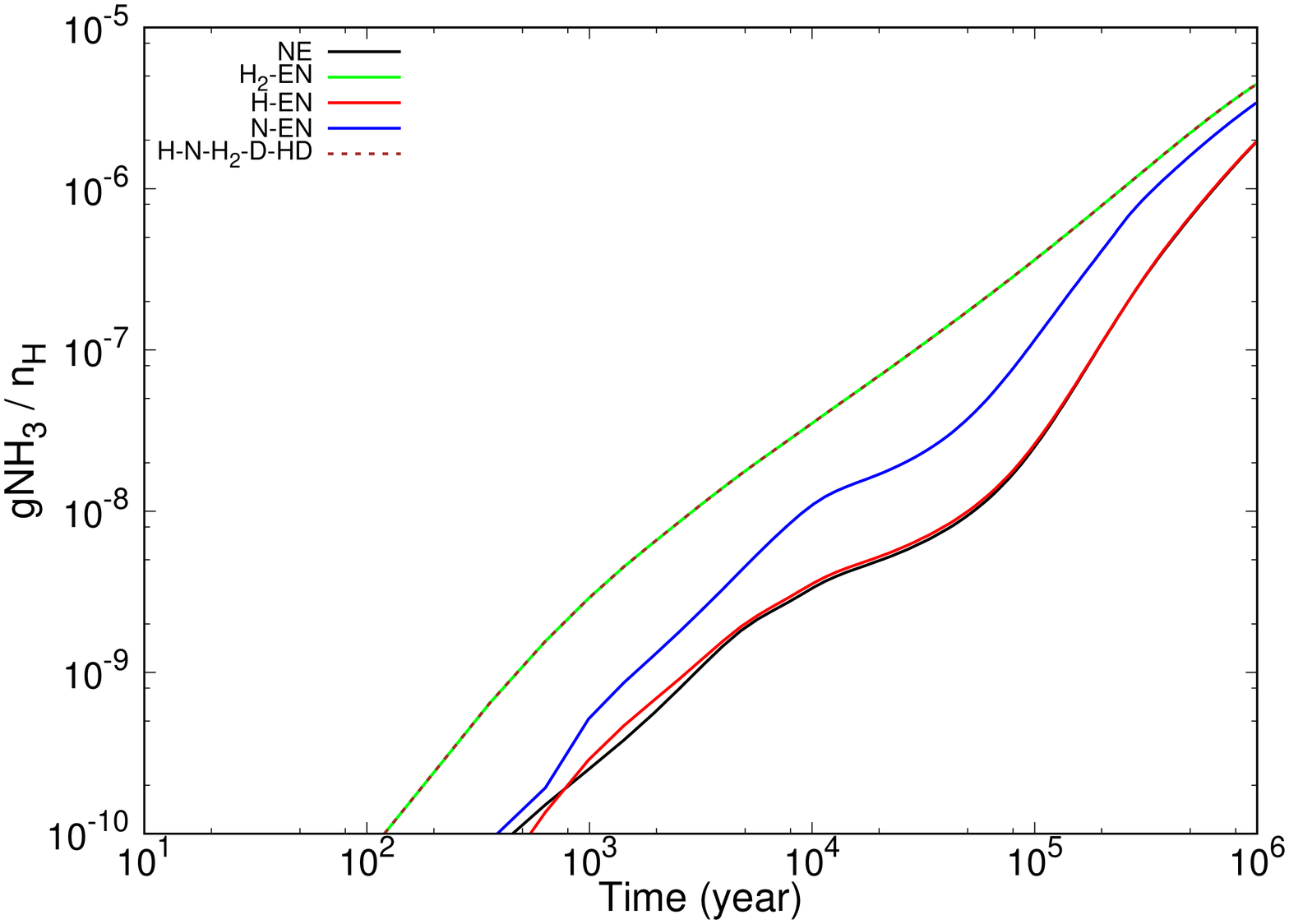}
    \caption{The time evolution of the abundances of ice phase water (first panel), methanol (second panel) and ammonia (third panel) is shown for n$_H=10^7$ cm$^{-3}$, $T=10$ K, and $R=0.35$. It shows a significant difference between the consideration of encounter desorption (solid green line for H$_2$, solid red line for H, and solid blue line for N) and without encounter desorption (black line).
The encounter desorption of H, N, H$_2$, D, and HD are collectively considered (brown dotted line) and show that it marginally deviates from the encounter desorption of H$_2$.}
    \label{fig:others}
\end{figure}

\begin{figure}
    \centering
    \includegraphics[height=8cm,width=7cm,angle=-90]{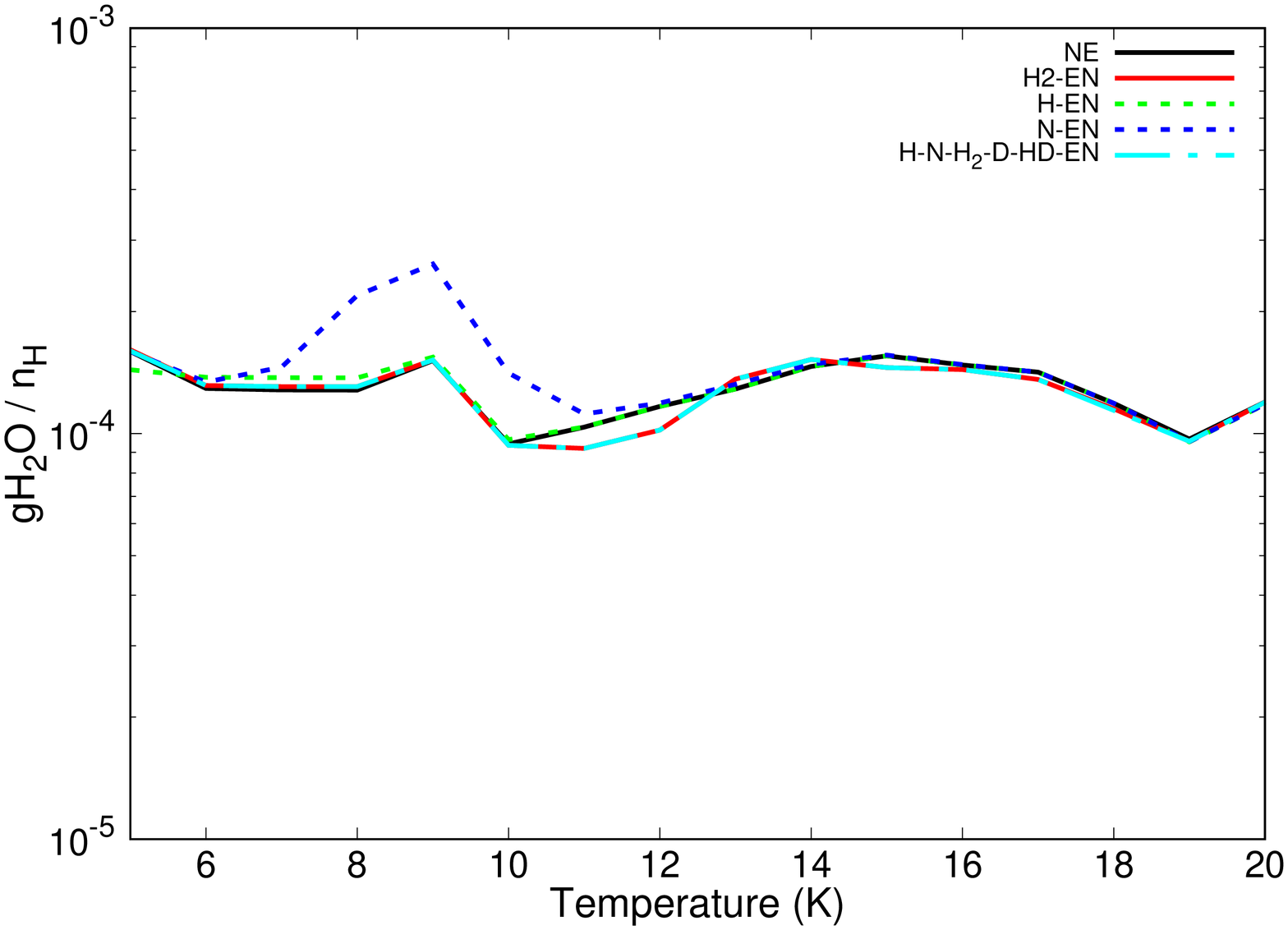}
    \includegraphics[height=8cm,width=7cm,angle=-90]{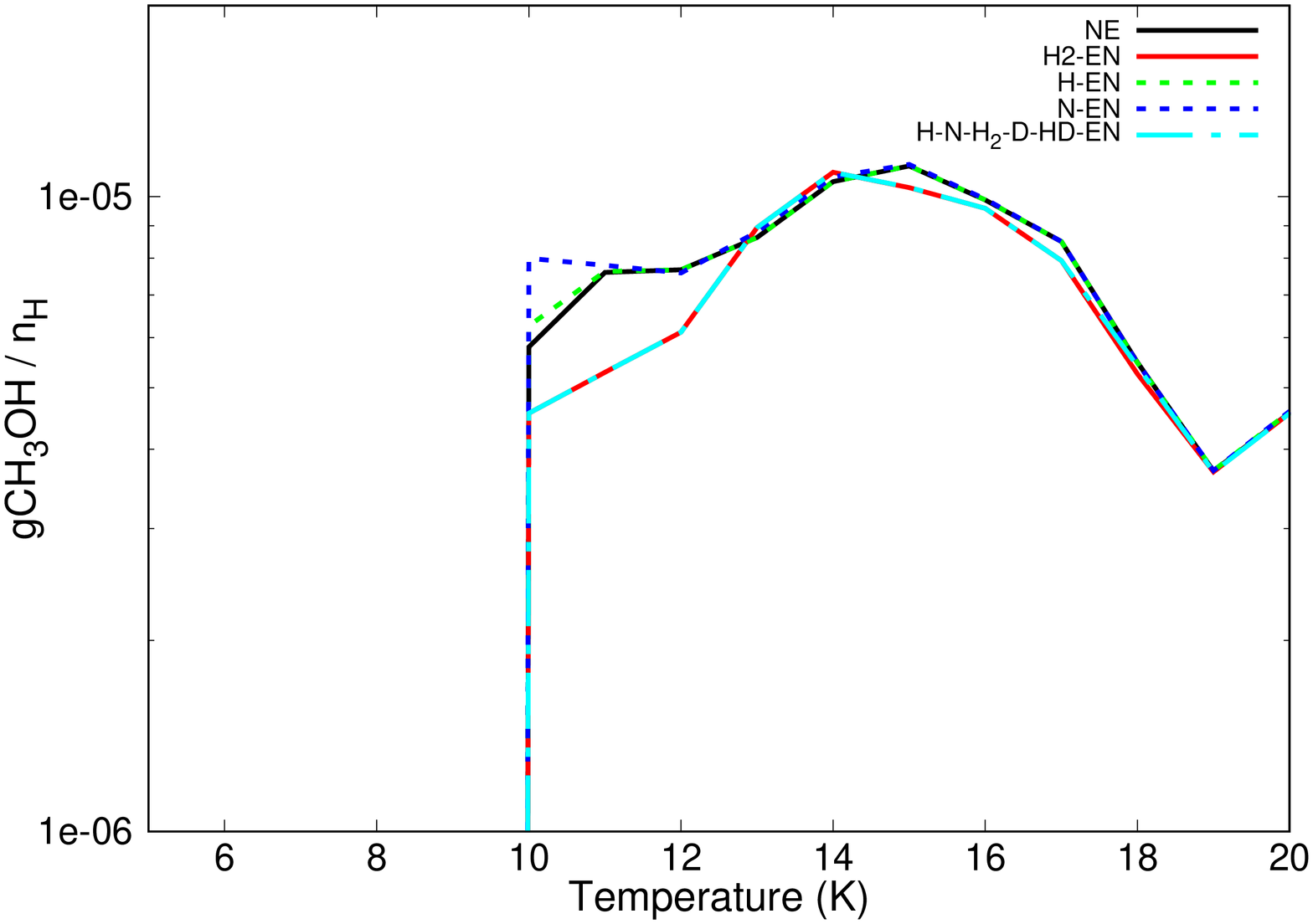}
    \includegraphics[height=8cm,width=7cm,angle=-90]{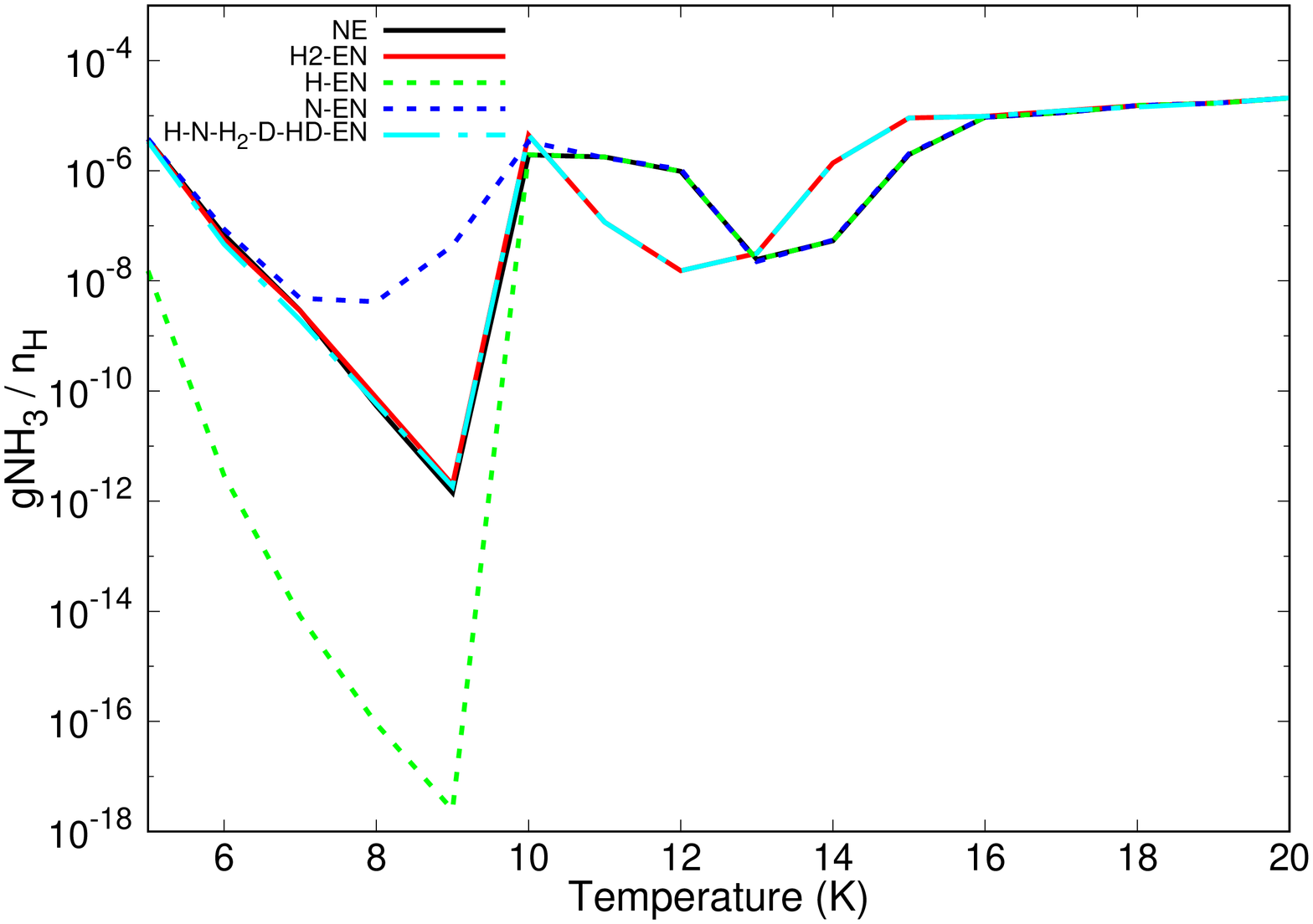}
    \caption{Temperature variation of the abundances of ice phase water (first panel), methanol (second panel), and ammonia (third panel) is shown for n$_H=10^7$ cm$^{-3}$ and $R=0.35$. It shows a significant difference between the consideration of encounter desorption and without encounter desorption (black line).
The encounter desorption of H, N, H$_2$, D, and HD are collectively considered, and as like Figure \ref{fig:others},  it marginally varies from the encounter desorption of H$_2$. 
}
    \label{fig:others-T}
\end{figure}

In Figure \ref{fig:abun}, we have shown the time evolution of the abundances of gH, gH$_2$, gN, gD, and gHD with n$_H=10^7$ cm$^{-3}$, $T=10$ K, and $R=0.35$. The encounter desorption of H$_2$ and without the encounter desorption effect are shown to show the differences. 
Figure \ref{fig:abun} depicts that the abundances of gN, gH, and gH$_2$, have a reasonably high surface coverage. Since these species have a reasonable hopping rate at the low temperature, encounter desorption of these species need to be considered in the chemical model. Here, we have included the encounter desorption of these species sequentially to check their effect on the final abundances of some of the key surface species (gH$_2$O, gCH$_3$OH, and gNH$_3$). To check the effect of encounter desorption of the other species, we have sequentially included the encounter desorption of H$_2$, H, and N. 
Figure \ref{fig:others} shows the time evolution of the encounter desorption of gH$_2$O, gCH$_3$OH, and gNH$_3$. We have already discussed the encounter desorption of gH$_2$ in section \ref{sec:H2-EN}.
Figure \ref{fig:others} shows that when we have included the encounter desorption of the H atom and N atom, the time evolution of the abundances shows significant changes in abundance. 
It depicts that considering the effect of encounter desorption of N atom can substantially increase the abundances of gH$_2$O, gCH$_3$OH, and gNH$_3$ for the physical condition considered here (n$_H=10^7$ cm$^{-3}$, $T=10$ K, and $R=0.35$). 
We further have included the encounter desorption of D and HD by considering the same BE as it was obtained for H and H$_2$ with the H$_2$ substrate. The cumulative effect (by considering the encounter desorption of H, H$_2$, N, D, and HD together) on the abundances is shown with the dotted curve. We have noticed that the abundance profile considering the cumulative effect shows a notable difference from that obtained with the no encounter desorption case. But the cumulative effect marginally differs from the encounter desorption effect of H$_2$. 
In Figure \ref{fig:others-T}, we have shown the temperature variation of the final abundances of water, methanol, and ammonia with respect to total hydrogen nuclei in all forms. It shows that the ice phase abundances of methanol, water, and ammonia can strongly deviate from the no encounter desorption case. As like the Figure \ref{fig:others}, we have also seen that the cumulative effect of the encounter desorption marginally deviates from the encounter desorption of H$_2$. Around $20$ K, we have noticed a great match between the cumulative encounter desorption case (dash-dotted cyan line), H$_2$ encounter desorption case (solid red line), and no encounter desorption case (solid black line). The right panel of Figure \ref{fig:H2-ratio} shows that as we have increased the temperature beyond $10$ K, the effect of the encounter desorption of H$_2$ starts to decrease. Around $20$ K, it roughly diminishes. Since the cumulative effect follows the nature of H$_2$ encounter desorption, it also matches with the no encounter desorption case at $\sim 20$ K.
 
\section{Conclusion} \label{sec:conclusions}
Here, we have provided realistic BEs of the interstellar species with the H$_2$ substrate. Supported with these BE values, we further have implemented our CMMC model to check the encounter desorption effect of H$_2$, H, and N on the interstellar ices. Following are the major highlights of this study.

\begin{itemize}

 \item  Our quantum chemical calculation finds a lower BE value ($\sim 10$ times) of all the species than it was obtained with the water \cite{das18} substrate. 

\item  Earlier in the literature, $\rm{E_D(H_2,H_2)=23}$ K \citep{cupp07,hinc15,chang20} and $\rm{E_D(H,H_2)=45}$ K \citep{cupp07,chang20} are used. Our quantum chemical calculations find an opposite trend with $\rm{E_D(H_2,H_2)=67}$ K and $\rm{E_D(H,H_2)=23}$ K. \cite{sil17} also explored that BE of the H$_2$ molecule always remains higher than that of the H atom considering different adsorbents like benzene, silica, and water cluster. The consideration of these updated adsorption energies show a significant deviation in the abundances of the surface species.

 \item  Our modeling results suggest that the inclusion of the encounter desorption of the H, H$_2$, and N  can affect the abundances of the major surface constituents like water, methanol, and ammonia. The cumulative effect roughly resembles a similar abundance with that obtained with the H$_2$'s encounter desorption only. For a bit higher temperature ($\sim 20$ K), when the encounter desorption effect of H$_2$ ceases, the encounter desorption of the cumulative cases exactly matches with the no encounter desorption case.

\end{itemize}

\section*{Data Availability Statement}
The original contributions presented in the study are included in the article/Supplementary Material; further inquiries can be directed to the corresponding author.

\section*{Conflict of Interest Statement}
The authors declare that the research was conducted in the
absence of any commercial or financial relationships that could be construed as a potential conflict of interest.

\section*{Author Contributions}
All authors listed have made a substantial, direct, and intellectual contribution to the work and approved it for publication.

\section*{Acknowledgments}
MS acknowledges DST, the Government of
India, for providing financial assistance through the DST-INSPIRE Fellowship [IF160109] scheme. SA and SS acknowledge Indian Centre for Space Physics for allowing them to continue their M.Sc. project work. This research was possible in part
due to a Grant-In-Aid from the Higher Education Department of the Government of West Bengal.

\section*{Supplementary Material}
The Supplementary Material for this article can be found online at:\\
{\tiny \url{https://www.frontiersin.org/articles/10.3389/fspas.2021.671622/full#supplementary-material}}

\clearpage
\bibliographystyle{frontiersinSCNS_ENG_HUMS} 
\bibliography{Encounter}

\begin{thebibliography}{47}
\providecommand{\natexlab}[1]{#1}
\expandafter\ifx\csname urlstyle\endcsname\relax
  \providecommand{\doi}[1]{doi:\discretionary{}{}{}#1}\else
  \providecommand{\doi}{doi:\discretionary{}{}{}\begingroup
  \urlstyle{rm}\Url}\fi
\providecommand{\selectlanguage}[1]{\relax}
\providecommand{\bibAnnoteFile}[1]{%
  \IfFileExists{#1}{\begin{quotation}\noindent\textsc{Key:} #1\\
  \textsc{Annotation:}\ \input{#1}\end{quotation}}{}}
\providecommand{\bibAnnote}[2]{%
  \begin{quotation}\noindent\textsc{Key:} #1\\
  \textsc{Annotation:}\ #2\end{quotation}}

\bibitem[{Bergin and Tafalla(2007)}]{berg07}
Bergin, E.~A. and Tafalla, M. (2007).
\newblock Cold dark clouds: The initial conditions for star formation.
\newblock \emph{Annual Review of Astronomy and Astrophysics} 45, 339--396.
\newblock \doi{10.1146/annurev.astro.45.071206.100404}
\bibAnnoteFile{berg07}

\bibitem[{{Boogert} et~al.(2015){Boogert}, {Gerakines}, and {Whittet}}]{boog15}
{Boogert}, A.~C.~A., {Gerakines}, P.~A., and {Whittet}, D. C.~B. (2015).
\newblock {Observations of the icy universe.}
\newblock \emph{Annual Review of Astronomy and Astrophysics} 53, 541--581.
\newblock \doi{10.1146/annurev-astro-082214-122348}
\bibAnnoteFile{boog15}

\bibitem[{{Canc{\`e}s} et~al.(1997){Canc{\`e}s}, {Mennucci}, and
  {Tomasi}}]{canc97}
{Canc{\`e}s}, E., {Mennucci}, B., and {Tomasi}, J. (1997).
\newblock {A new integral equation formalism for the polarizable continuum
  model: Theoretical background and applications to isotropic and anisotropic
  dielectrics}.
\newblock \emph{Journal of Chemical Physics} 107, 3032--3041.
\newblock \doi{10.1063/1.474659}
\bibAnnoteFile{canc97}

\bibitem[{{Chaabouni} et~al.(2012){Chaabouni}, {Bergeron}, {Baouche}, {Dulieu},
  {Matar}, {Congiu} et~al.}]{chaa12}
{Chaabouni}, H., {Bergeron}, H., {Baouche}, S., {Dulieu}, F., {Matar}, E.,
  {Congiu}, E., et~al. (2012).
\newblock {Sticking coefficient of hydrogen and deuterium on silicates under
  interstellar conditions}.
\newblock \emph{Astronomy and Astrophysics} 538, A128.
\newblock \doi{10.1051/0004-6361/201117409}
\bibAnnoteFile{chaa12}

\bibitem[{{Chakrabarti} et~al.(2006a){Chakrabarti}, {Das}, {Acharyya}, and
  {Chakrabarti}}]{chak06a}
{Chakrabarti}, S.~K., {Das}, A., {Acharyya}, K., and {Chakrabarti}, S. (2006a).
\newblock {Effective grain surface area in the formation of molecular hydrogen
  in interstellar clouds}.
\newblock \emph{Astronomy and Astrophysics} 457, 167--170.
\newblock \doi{10.1051/0004-6361:20065335}
\bibAnnoteFile{chak06a}

\bibitem[{{Chakrabarti} et~al.(2006b){Chakrabarti}, {Das}, {Acharyya}, and
  {Chakrabarti}}]{chak06b}
{Chakrabarti}, S.~K., {Das}, A., {Acharyya}, K., and {Chakrabarti}, S. (2006b).
\newblock {Recombination efficiency of molecular hydrogen on interstellar
  grains-II. A numerical study}.
\newblock \emph{Bulletin of the Astronomical Society of India} 34, 299
\bibAnnoteFile{chak06b}

\bibitem[{{Chang} et~al.(2021){Chang}, {Zheng}, {Zhang}, {Quan}, {Lu}, {Meng}
  et~al.}]{chang20}
{Chang}, Q., {Zheng}, X.-L., {Zhang}, X., {Quan}, D.-H., {Lu}, Y., {Meng},
  Q.-K., et~al. (2021).
\newblock {On the encounter desorption of hydrogen atoms on an ice mantle}.
\newblock \emph{Research in Astronomy and Astrophysics} 21, 039.
\newblock \doi{10.1088/1674-4527/21/2/39}
\bibAnnoteFile{chang20}

\bibitem[{{Collings} et~al.(2004){Collings}, {Anderson}, {Chen}, {Dever},
  {Viti}, {Williams} et~al.}]{coll04}
{Collings}, M.~P., {Anderson}, M.~A., {Chen}, R., {Dever}, J.~W., {Viti}, S.,
  {Williams}, D.~A., et~al. (2004).
\newblock {A laboratory survey of the thermal desorption of astrophysically
  relevant molecules}.
\newblock \emph{Monthly Notices of the Royal Astronomical Society} 354,
  1133--1140.
\newblock \doi{10.1111/j.1365-2966.2004.08272.x}
\bibAnnoteFile{coll04}

\bibitem[{{Cuppen} and {Herbst}(2007)}]{cupp07}
{Cuppen}, H.~M. and {Herbst}, E. (2007).
\newblock {Simulation of the Formation and Morphology of Ice Mantles on
  Interstellar Grains}.
\newblock \emph{The Astrophysical Journal} 668, 294--309.
\newblock \doi{10.1086/521014}
\bibAnnoteFile{cupp07}

\bibitem[{{Das} et~al.(2008a){Das}, {Acharyya}, {Chakrabarti}, and
  {Chakrabarti}}]{das08a}
{Das}, A., {Acharyya}, K., {Chakrabarti}, S., and {Chakrabarti}, S.~K. (2008a).
\newblock {Formation of water and methanol in star forming molecular clouds}.
\newblock \emph{Astronomy and Astrophysics} 486, 209--220.
\newblock \doi{10.1051/0004-6361:20078422}
\bibAnnoteFile{das08a}

\bibitem[{{Das} et~al.(2010){Das}, {Acharyya}, and {Chakrabarti}}]{das10}
{Das}, A., {Acharyya}, K., and {Chakrabarti}, S.~K. (2010).
\newblock {Effects of initial condition and cloud density on the composition of
  the grain mantle}.
\newblock \emph{Monthly Notices of the Royal Astronomical Society} 409,
  789--800.
\newblock \doi{10.1111/j.1365-2966.2010.17343.x}
\bibAnnoteFile{das10}

\bibitem[{{Das} and {Chakrabarti}(2011)}]{das11}
{Das}, A. and {Chakrabarti}, S.~K. (2011).
\newblock {Composition and evolution of interstellar grain mantle under the
  effects of photodissociation}.
\newblock \emph{Monthly Notices of the Royal Astronomical Society} 418,
  545--555.
\newblock \doi{10.1111/j.1365-2966.2011.19503.x}
\bibAnnoteFile{das11}

\bibitem[{{Das} et~al.(2015b){Das}, {Majumdar}, {Chakrabarti}, and
  {Sahu}}]{das15b}
{Das}, A., {Majumdar}, L., {Chakrabarti}, S.~K., and {Sahu}, D. (2015b).
\newblock {Deuterium enrichment of the interstellar medium}.
\newblock \emph{New Astronomy} 35, 53--70.
\newblock \doi{10.1016/j.newast.2014.07.006}
\bibAnnoteFile{das15b}

\bibitem[{{Das} et~al.(2015a){Das}, {Majumdar}, {Sahu}, {Gorai}, {Sivaraman},
  and {Chakrabarti}}]{das15a}
{Das}, A., {Majumdar}, L., {Sahu}, D., {Gorai}, P., {Sivaraman}, B., and
  {Chakrabarti}, S.~K. (2015a).
\newblock {Methyl Acetate and Its Singly Deuterated Isotopomers in the
  Interstellar Medium}.
\newblock \emph{The Astrophysical Journal} 808, 21.
\newblock \doi{10.1088/0004-637X/808/1/21}
\bibAnnoteFile{das15a}

\bibitem[{{Das} et~al.(2016){Das}, {Sahu}, {Majumdar}, and
  {Chakrabarti}}]{das16}
{Das}, A., {Sahu}, D., {Majumdar}, L., and {Chakrabarti}, S.~K. (2016).
\newblock {Deuterium enrichment of the interstellar grain mantle}.
\newblock \emph{Monthly Notices of the Royal Astronomical Society} 455,
  540--551.
\newblock \doi{10.1093/mnras/stv2264}
\bibAnnoteFile{das16}

\bibitem[{{Das} et~al.(2018){Das}, {Sil}, {Gorai}, {Chakrabarti}, and
  {Loison}}]{das18}
{Das}, A., {Sil}, M., {Gorai}, P., {Chakrabarti}, S. i.~K., and {Loison}, J.~C.
  (2018).
\newblock {An Approach to Estimate the Binding Energy of Interstellar Species}.
\newblock \emph{The Astrophysical Journal Supplement Series} 237, 9.
\newblock \doi{10.3847/1538-4365/aac886}
\bibAnnoteFile{das18}

\bibitem[{{Dulieu} et~al.(2013){Dulieu}, {Congiu}, {Noble}, {Baouche},
  {Chaabouni}, {Moudens} et~al.}]{duli13}
{Dulieu}, F., {Congiu}, E., {Noble}, J., {Baouche}, S., {Chaabouni}, H.,
  {Moudens}, A., et~al. (2013).
\newblock {How micron-sized dust particles determine the chemistry of our
  Universe}.
\newblock \emph{Scientific Reports} 3, 1338.
\newblock \doi{10.1038/srep01338}
\bibAnnoteFile{duli13}

\bibitem[{{Dunning}(1989)}]{dunn89}
{Dunning}, J., Thom~H. (1989).
\newblock {Gaussian basis sets for use in correlated molecular calculations. I.
  The atoms boron through neon and hydrogen}.
\newblock \emph{The Journal of Chemical Physics} 90, 1007--1023.
\newblock \doi{10.1063/1.456153}
\bibAnnoteFile{dunn89}

\bibitem[{Frisch et~al.(2013)Frisch, Trucks, Schlegel, Scuseria, Robb,
  Cheeseman et~al.}]{fris13}
[Dataset] Frisch, M.~J., Trucks, G.~W., Schlegel, H.~B., Scuseria, G.~E., Robb,
  M.~A., Cheeseman, J.~R., et~al. (2013).
\newblock Gaussian 09 {R}evision {D}.01.
\newblock Gaussian Inc. Wallingford CT
\bibAnnoteFile{fris13}

\bibitem[{{Garrod} and {Pauly}(2011)}]{garr11}
{Garrod}, R.~T. and {Pauly}, T. (2011).
\newblock {On the Formation of CO$_{2}$ and Other Interstellar Ices}.
\newblock \emph{The Astrophysical Journal} 735, 15.
\newblock \doi{10.1088/0004-637X/735/1/15}
\bibAnnoteFile{garr11}

\bibitem[{{Garrod} et~al.(2007){Garrod}, {Wakelam}, and {Herbst}}]{garr07}
{Garrod}, R.~T., {Wakelam}, V., and {Herbst}, E. (2007).
\newblock {Non-thermal desorption from interstellar dust grains via exothermic
  surface reactions}.
\newblock \emph{Astronomy and Astrophysics} 467, 1103--1115.
\newblock \doi{10.1051/0004-6361:20066704}
\bibAnnoteFile{garr07}

\bibitem[{{Gibb} et~al.(2004){Gibb}, {Whittet}, {Boogert}, and
  {Tielens}}]{gibb04}
{Gibb}, E.~L., {Whittet}, D.~C.~B., {Boogert}, A.~C.~A., and {Tielens},
  A.~G.~G.~M. (2004).
\newblock {Interstellar Ice: The Infrared Space Observatory Legacy}.
\newblock \emph{The Astrophysical Journal Supplement Series} 151, 35--73.
\newblock \doi{10.1086/381182}
\bibAnnoteFile{gibb04}

\bibitem[{Gibb et~al.(2000)Gibb, Whittet, Schutte, Boogert, Chiar, Ehrenfreund
  et~al.}]{gibb00}
Gibb, E.~L., Whittet, D. C.~B., Schutte, W.~A., Boogert, A. C.~A., Chiar,
  J.~E., Ehrenfreund, P., et~al. (2000).
\newblock An inventory of interstellar ices toward the embedded protostar w33a.
\newblock \emph{The Astrophysical Journal} 536, 347--356.
\newblock \doi{10.1086/308940}
\bibAnnoteFile{gibb00}

\bibitem[{{Gorai} et~al.(2020){Gorai}, {Bhat}, {Sil}, {Mondal}, {Ghosh},
  {Chakrabarti} et~al.}]{gora20b}
{Gorai}, P., {Bhat}, B., {Sil}, M., {Mondal}, S.~K., {Ghosh}, R.,
  {Chakrabarti}, S.~K., et~al. (2020).
\newblock {Identification of Prebiotic Molecules Containing Peptide-like Bonds
  in a Hot Molecular Core, G10.47+0.03}.
\newblock \emph{The Astrophysical Journal}
\bibAnnoteFile{gora20b}

\bibitem[{{Gorai} et~al.(2017a){Gorai}, {Das}, {Das}, {Sivaraman}, {Etim}, and
  {Chakrabarti}}]{gora17a}
{Gorai}, P., {Das}, A., {Das}, A., {Sivaraman}, B., {Etim}, E.~E., and
  {Chakrabarti}, S. i.~K. (2017a).
\newblock {A Search for Interstellar Monohydric Thiols}.
\newblock \emph{The Astrophysical Journal} 836, 70.
\newblock \doi{10.3847/1538-4357/836/1/70}
\bibAnnoteFile{gora17a}

\bibitem[{{Gorai} et~al.(2017b){Gorai}, {Das}, {Majumdar}, {Chakrabarti},
  {Sivaraman}, and {Herbst}}]{gora17b}
{Gorai}, P., {Das}, A., {Majumdar}, L., {Chakrabarti}, S.~K., {Sivaraman}, B.,
  and {Herbst}, E. (2017b).
\newblock {The Possibility of Forming Propargyl Alcohol in the Interstellar
  Medium}.
\newblock \emph{Molecular Astrophysics} 6, 36--46.
\newblock \doi{10.1016/j.molap.2017.01.004}
\bibAnnoteFile{gora17b}

\bibitem[{Gorai et~al.(2020a)Gorai, Sil, Das, Sivaraman, Chakrabarti, Ioppolo
  et~al.}]{gora20a}
Gorai, P., Sil, M., Das, A., Sivaraman, B., Chakrabarti, S.~K., Ioppolo, S.,
  et~al. (2020a).
\newblock Systematic study on the absorption features of interstellar ices in
  the presence of impurities.
\newblock \emph{ACS Earth and Space Chemistry} 4, 920–946.
\newblock \doi{10.1021/acsearthspacechem.0c00098}
\bibAnnoteFile{gora20a}

\bibitem[{{Hasegawa} et~al.(1992){Hasegawa}, {Herbst}, and {Leung}}]{hase92}
{Hasegawa}, T.~I., {Herbst}, E., and {Leung}, C.~M. (1992).
\newblock {Models of Gas-Grain Chemistry in Dense Interstellar Clouds with
  Complex Organic Molecules}.
\newblock \emph{The Astrophysical Journal Supplement Series} 82, 167.
\newblock \doi{10.1086/191713}
\bibAnnoteFile{hase92}

\bibitem[{{Hincelin} et~al.(2015){Hincelin}, {Chang}, and {Herbst}}]{hinc15}
{Hincelin}, U., {Chang}, Q., and {Herbst}, E. (2015).
\newblock {A new and simple approach to determine the abundance of hydrogen
  molecules on interstellar ice mantles}.
\newblock \emph{Astronomy and Astrophysics} 574, A24.
\newblock \doi{10.1051/0004-6361/201424807}
\bibAnnoteFile{hinc15}

\bibitem[{{Keto} and {Caselli}(2008)}]{keto08}
{Keto}, E. and {Caselli}, P. (2008).
\newblock {The Different Structures of the Two Classes of Starless Cores}.
\newblock \emph{The Astrophysical Journal} 683, 238--247.
\newblock \doi{10.1086/589147}
\bibAnnoteFile{keto08}

\bibitem[{{Li}(2004)}]{li04}
{Li}, A. (2004).
\newblock {Interaction of Nanoparticles with Radiation}.
\newblock In \emph{Astrophysics of Dust}, eds. A.~N. {Witt}, G.~C. {Clayton},
  and B.~T. {Draine}. vol. 309 of \emph{Astronomical Society of the Pacific
  Conference Series}, 417
\bibAnnoteFile{li04}

\bibitem[{{McElroy} et~al.(2013){McElroy}, {Walsh}, {Markwick}, {Cordiner},
  {Smith}, and {Millar}}]{mcel13}
{McElroy}, D., {Walsh}, C., {Markwick}, A.~J., {Cordiner}, M.~A., {Smith}, K.,
  and {Millar}, T.~J. (2013).
\newblock {The UMIST database for astrochemistry 2012}.
\newblock \emph{Astronomy and Astrophysics} 550, A36.
\newblock \doi{10.1051/0004-6361/201220465}
\bibAnnoteFile{mcel13}

\bibitem[{{Noble} et~al.(2012){Noble}, {Congiu}, {Dulieu}, and
  {Fraser}}]{nobl12}
{Noble}, J.~A., {Congiu}, E., {Dulieu}, F., and {Fraser}, H.~J. (2012).
\newblock {Thermal desorption characteristics of CO, O$_{2}$ and CO$_{2}$ on
  non-porous water, crystalline water and silicate surfaces at submonolayer and
  multilayer coverages}.
\newblock \emph{Monthly Notices of the Royal Astronomical Society} 421,
  768--779.
\newblock \doi{10.1111/j.1365-2966.2011.20351.x}
\bibAnnoteFile{nobl12}

\bibitem[{Pagani et~al.(2013)Pagani, Lesaffre, Roueff, Jorfi, Honvault,
  González-Lezana et~al.}]{paga12}
Pagani, L., Lesaffre, P., Roueff, E., Jorfi, M., Honvault, P.,
  González-Lezana, T., et~al. (2013).
\newblock \emph{Philosophical Transactions of the Royal Society A:
  Mathematical, Physical and Engineering Sciences} 370, 5201--5212.
\newblock \doi{11.1098/rsta.2012.0027}
\bibAnnoteFile{paga12}

\bibitem[{{Roberts} and {Millar}(2000)}]{robe00}
{Roberts}, H. and {Millar}, T.~J. (2000).
\newblock {Modelling of deuterium chemistry and its application to molecular
  clouds}.
\newblock \emph{Astronomy and Astrophysics} 361, 388--398
\bibAnnoteFile{robe00}

\bibitem[{{Rohatgi}(2020)}]{roha20}
[Dataset] {Rohatgi}, A. (2020).
\newblock {Webplotdigitizer: Version 4.4}
\bibAnnoteFile{roha20}

\bibitem[{{Ruaud} et~al.(2015){Ruaud}, {Loison}, {Hickson}, {Gratier},
  {Hersant}, and {Wakelam}}]{ruau15}
{Ruaud}, M., {Loison}, J.~C., {Hickson}, K.~M., {Gratier}, P., {Hersant}, F.,
  and {Wakelam}, V. (2015).
\newblock {Modelling complex organic molecules in dense regions: Eley-Rideal
  and complex induced reaction}.
\newblock \emph{Monthly Notices of the Royal Astronomical Society} 447,
  4004--4017.
\newblock \doi{10.1093/mnras/stu2709}
\bibAnnoteFile{ruau15}

\bibitem[{{Sandford} and {Allamandola}(1993)}]{sand93}
{Sandford}, S.~A. and {Allamandola}, L.~J. (1993).
\newblock {H2 in Interstellar and Extragalactic Ices: Infrared Characteristics,
  Ultraviolet Production, and Implications}.
\newblock \emph{The Astrophysical Journal Letters} 409, L65.
\newblock \doi{10.1086/186861}
\bibAnnoteFile{sand93}

\bibitem[{{Semenov} et~al.(2010){Semenov}, {Hersant}, {Wakelam}, {Dutrey},
  {Chapillon}, {Guilloteau} et~al.}]{seme10}
{Semenov}, D., {Hersant}, F., {Wakelam}, V., {Dutrey}, A., {Chapillon}, E.,
  {Guilloteau}, S., et~al. (2010).
\newblock {Chemistry in disks. IV. Benchmarking gas-grain chemical models with
  surface reactions}.
\newblock \emph{Astronomy and Astrophysics} 522, A42.
\newblock \doi{10.1051/0004-6361/201015149}
\bibAnnoteFile{seme10}

\bibitem[{{Sil} et~al.(2018){Sil}, {Gorai}, {Das}, {Bhat}, {Etim}, and
  {Chakrabarti}}]{sil18}
{Sil}, M., {Gorai}, P., {Das}, A., {Bhat}, B., {Etim}, E.~E., and
  {Chakrabarti}, S.~K. (2018).
\newblock {Chemical Modeling for Predicting the Abundances of Certain Aldimines
  and Amines in Hot Cores}.
\newblock \emph{The Astrophysical Journal} 853, 139.
\newblock \doi{10.3847/1538-4357/aa984d}
\bibAnnoteFile{sil18}

\bibitem[{{Sil} et~al.(2017){Sil}, {Gorai}, {Das}, {Sahu}, and
  {Chakrabarti}}]{sil17}
{Sil}, M., {Gorai}, P., {Das}, A., {Sahu}, D., and {Chakrabarti}, S.~K. (2017).
\newblock {Adsorption energies of H and H$_{2}$: a quantum-chemical study}.
\newblock \emph{European Physical Journal D} 71, 45.
\newblock \doi{10.1140/epjd/e2017-70610-4}
\bibAnnoteFile{sil17}

\bibitem[{Tomasi et~al.(2005)Tomasi, Mennucci, and Cammi}]{toma05}
Tomasi, J., Mennucci, B., and Cammi, R. (2005).
\newblock Quantum mechanical continuum solvation models.
\newblock \emph{Chemical Reviews} 105, 2999--3094.
\newblock \doi{10.1021/cr9904009}.
\newblock PMID: 16092826
\bibAnnoteFile{toma05}

\bibitem[{{Vidali} et~al.(1991){Vidali}, {Ihm}, {Kim}, and {Cole}}]{vida91}
{Vidali}, G., {Ihm}, G., {Kim}, H.-Y., and {Cole}, M.~W. (1991).
\newblock {Potentials of physical adsorption}.
\newblock \emph{Surface Science Reports} 12, 135--181.
\newblock \doi{10.1016/0167-5729(91)90012-M}
\bibAnnoteFile{vida91}

\bibitem[{{Wakelam} et~al.(2017){Wakelam}, {Loison}, {Mereau}, and
  {Ruaud}}]{wake17}
{Wakelam}, V., {Loison}, J.~C., {Mereau}, R., and {Ruaud}, M. (2017).
\newblock {Binding energies: New values and impact on the efficiency of
  chemical desorption}.
\newblock \emph{Molecular Astrophysics} 6, 22--35.
\newblock \doi{10.1016/j.molap.2017.01.002}
\bibAnnoteFile{wake17}

\bibitem[{{Ward} et~al.(2012){Ward}, {Hogg}, and {Price}}]{ward12}
{Ward}, M.~D., {Hogg}, I.~A., and {Price}, S.~D. (2012).
\newblock {Thermal reactions of oxygen atoms with CS$_{2}$ at low temperatures
  on interstellar dust}.
\newblock \emph{Monthly Notices of the Royal Astronomical Society} 425,
  1264--1269.
\newblock \doi{10.1111/j.1365-2966.2012.21520.x}
\bibAnnoteFile{ward12}

\bibitem[{{Whittet} et~al.(2007){Whittet}, {Shenoy}, {Bergin}, {Chiar},
  {Gerakines}, {Gibb} et~al.}]{whit07}
{Whittet}, D.~C.~B., {Shenoy}, S.~S., {Bergin}, E.~A., {Chiar}, J.~E.,
  {Gerakines}, P.~A., {Gibb}, E.~L., et~al. (2007).
\newblock {The Abundance of Carbon Dioxide Ice in the Quiescent Intracloud
  Medium}.
\newblock \emph{The Astrophysical Journal} 655, 332--341.
\newblock \doi{10.1086/509772}
\bibAnnoteFile{whit07}

\bibitem[{Öberg et~al.(2008)Öberg, Boogert, Pontoppidan, Blake, Evans, Lahuis
  et~al.}]{ober08}
Öberg, K.~I., Boogert, A.~C.~A., Pontoppidan, K.~M., Blake, G.~A., Evans,
  N.~J., Lahuis, F., et~al. (2008).
\newblock The c2dspitzerspectroscopic survey of ices around low-mass young
  stellar objects. {III}. {CH}4.
\newblock \emph{The Astrophysical Journal} 678, 1032--1041.
\newblock \doi{10.1086/533432}
\bibAnnoteFile{ober08}

\end{thebibliography}

\end{document}